\documentclass[twoside,twocolumn,9pt]{article}
\usepackage{extsizes}
\usepackage[super,sort&compress,comma]{natbib}
\usepackage[version=3]{mhchem}
\usepackage[left=1.5cm, right=1.5cm, top=1.785cm, bottom=2.0cm]{geometry}
\usepackage{balance}
\usepackage{times,mathptmx}
\usepackage{sectsty}
\usepackage{graphicx}
\usepackage{lastpage}
\usepackage[format=plain,justification=raggedright,singlelinecheck=false,font={stretch=1.125,small,sf},labelfont=bf,labelsep=space]{caption}
\usepackage{float}
\usepackage{fancyhdr}
\usepackage{fnpos}
\usepackage[english]{babel}
\usepackage{array}
\usepackage{droidsans}
\usepackage{charter}
\usepackage[T1]{fontenc}
\usepackage[usenames,dvipsnames]{xcolor}
\usepackage{setspace}
\usepackage[compact]{titlesec}
\addto{\captionsenglish}{%
  
}

\usepackage{comment}
\usepackage{array}
\usepackage{droidsans}
\usepackage{charter}
\usepackage[T1]{fontenc}
\usepackage[usenames,dvipsnames]{xcolor}
\usepackage{setspace}
\usepackage[compact]{titlesec}
\usepackage{hyperref}
\usepackage{stackengine}
\usepackage{bm}
\usepackage{siunitx}
\usepackage{physics}
\usepackage{subcaption}
\usepackage{latexsym}
\usepackage{amsmath}
\usepackage{amsfonts}
\usepackage{amssymb}
\usepackage{amsbsy}
\usepackage{wasysym}
\usepackage{marvosym}
\usepackage{oplotsymbl}

\newcommand{\onlinecite}[1]{\hspace{-1 ex} \nocite{#1}\citenum{#1}}

\definecolor{cream}{RGB}{222,217,201}

\graphicspath{{./graphics/},{./head_foot.}}

\begin{document}

\pagestyle{fancy}
\thispagestyle{plain}
\fancypagestyle{plain}{
%%%HEADER%%%
\renewcommand{\headrulewidth}{0pt}
}
%%%END OF HEADER%%%

%%%PAGE SETUP - Please do not change any commands within this section%%%
\makeFNbottom
\makeatletter
\renewcommand\LARGE{\@setfontsize\LARGE{15pt}{17}}
\renewcommand\Large{\@setfontsize\Large{12pt}{14}}
\renewcommand\large{\@setfontsize\large{10pt}{12}}
\renewcommand\footnotesize{\@setfontsize\footnotesize{7pt}{10}}
\makeatother

\renewcommand{\thefootnote}{\fnsymbol{footnote}}
\renewcommand\footnoterule{\vspace*{1pt}% 
\color{cream}\hrule width 3.5in height 0.4pt \color{black}\vspace*{5pt}} 
\setcounter{secnumdepth}{5}

\makeatletter 
\renewcommand\@biblabel[1]{#1}            
\renewcommand\@makefntext[1]% 
{\noindent\makebox[0pt][r]{\@thefnmark\,}#1}
\makeatother 
\renewcommand{\figurename}{\small{Fig.}~}
\sectionfont{\sffamily\Large}
\subsectionfont{\normalsize}
\subsubsectionfont{\bf}
\setstretch{1.125} %In particular, please do not alter this line.
\setlength{\skip\footins}{0.8cm}
\setlength{\footnotesep}{0.25cm}
\setlength{\jot}{10pt}
\titlespacing*{\section}{0pt}{4pt}{4pt}
\titlespacing*{\subsection}{0pt}{15pt}{1pt}
%%%END OF PAGE SETUP%%%

%%%FOOTER%%%
\fancyfoot{}
\fancyfoot[LO,RE]{\vspace{-7.1pt}\includegraphics[height=9pt]{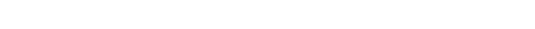}}
\fancyfoot[CO]{\vspace{-7.1pt}\hspace{13.2cm}\includegraphics{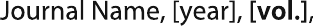}}
\fancyfoot[CE]{\vspace{-7.2pt}\hspace{-14.2cm}\includegraphics{RF}}
\fancyfoot[RO]{\footnotesize{\sffamily{1--\pageref{LastPage} ~\textbar  \hspace{2pt}\thepage}}}
\fancyfoot[LE]{\footnotesize{\sffamily{\thepage~\textbar\hspace{3.45cm} 1--\pageref{LastPage}}}}
\fancyhead{}
\renewcommand{\headrulewidth}{0pt} 
\renewcommand{\footrulewidth}{0pt}
\setlength{\arrayrulewidth}{1pt}
\setlength{\columnsep}{6.5mm}
\setlength\bibsep{1pt}
%%%END OF FOOTER%%%

%%%FIGURE SETUP - please do not change any commands within this section%%%
\makeatletter 
\newlength{\figrulesep} 
\setlength{\figrulesep}{0.5\textfloatsep} 

\newcommand{\topfigrule}{\vspace*{-1pt}% 
\noindent{\color{cream}\rule[-\figrulesep]{\columnwidth}{1.5pt}} }

\newcommand{\botfigrule}{\vspace*{-2pt}% 
\noindent{\color{cream}\rule[\figrulesep]{\columnwidth}{1.5pt}} }

\newcommand{\dblfigrule}{\vspace*{-1pt}% 
\noindent{\color{cream}\rule[-\figrulesep]{\textwidth}{1.5pt}} }

\makeatother
%%%END OF FIGURE SETUP%%%

%%%TITLE, AUTHORS AND ABSTRACT%%%
\twocolumn[
  \begin{@twocolumnfalse}
{\includegraphics[height=30pt]{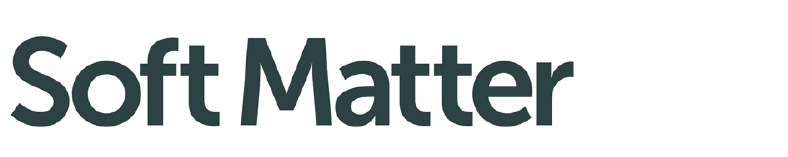}\hfill\raisebox{0pt}[0pt][0pt]{\includegraphics[height=55pt]{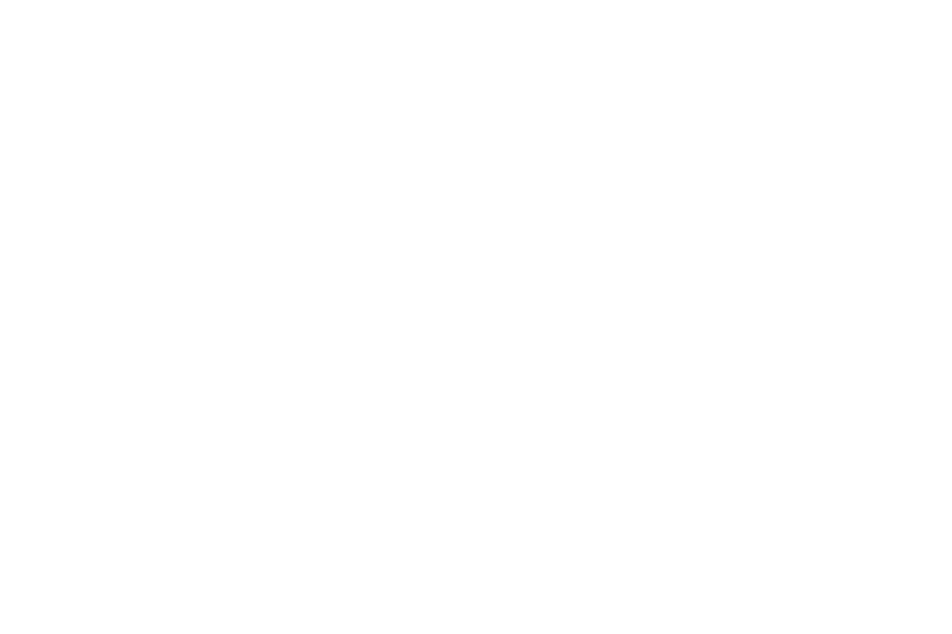}}\\[1ex]
\includegraphics[width=18.5cm]{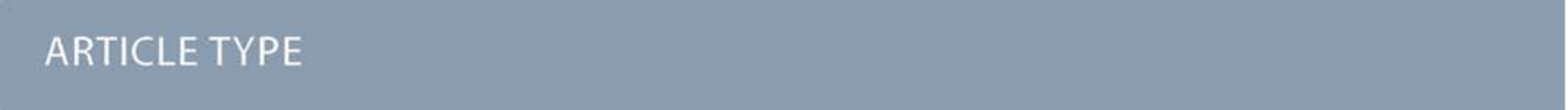}}\par
\vspace{1em}
\sffamily
\begin{tabular}{m{4.5cm} p{13.5cm} }

\includegraphics{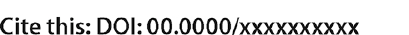} & \noindent\LARGE{\textbf{Emergent collective behavior of active Brownian particles by visual perception$^\dag$}} \\%Article title goes here instead of the text "This is the title"
\vspace{0.3cm} & \vspace{0.3cm} \\

 & \noindent\large{Rajendra Singh Negi,\textit{$^{a}$} }  Roland G. Winkler, and Gerhard Gompper,\textit{$^{a\ddag}$ } \\%Author names go here instead of "Full name", etc.

\includegraphics{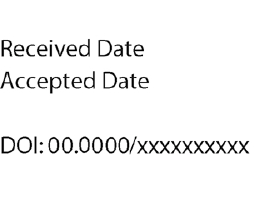} & \noindent\normalsize{Systems comprised of self-steering active Brownian particles are studied via simulations for a minimal cognitive flocking model. The dynamics of the active Brownian particles is extended by an orientational response with limited maneuverability to an instantaneous visual input of the positions of neighbors within a vision cone and a cut-off radius.  The system exhibits large-scale self-organized structures, which depend on selected parameter values, and, in particular, the presence of excluded-volume interactions.  The emergent structures in two dimensions, such as worms, worm-aggregate coexistence, and hexagonally close-packed structures, are analysed and phase diagrams are constructed. The analysis of the particle's mean-square displacement shows ABP-like dynamics for dilute systems and the worm phase. In the limit of densely packed structures, the active diffusion coefficient is significantly smaller and depends on the number of particles in the cluster. Our analysis of the cluster-growth dynamics shows distinct differences to processes in systems of short-range attractive colloids in equilibrium. Specifically, the characteristic time for the growth and decay of clusters of a particular size is longer than that of isotropically attractive colloids, which we attribute to the non-reciprocal nature of the directed visual perception. Our simulations reveal a strong interplay between ABP-characteristic interactions, such as volume exclusion and rotational diffusion, and cognitive-based interactions and navigation. 
} \\%The abstrast goes here instead of the text "The abstract should be..."

\end{tabular}

 \end{@twocolumnfalse} \vspace{0.6cm}
]
%%%END OF TITLE, AUTHORS AND ABSTRACT%%%

%%%FONT SETUP - please do not change any commands within this section
\renewcommand*\rmdefault{bch}\normalfont\upshape
\rmfamily
\section*{}
\vspace{-1cm}

%%%FOOTNOTES%%%

\footnotetext{\textit{$^{a}$~ Theoretical Physics of Living Matter, Institute of Biological Information Processing and Institute for Advanced Simulation, Forschungszentrum J{\"u}lich, 52428 J{\"u}lich, Germany}}
\footnotetext{\textit{$^{\ddag}$~g.gomper@fz-juelich.de} \\
$\dag$ Electronic supplementary information (ESI) available. See DOI: ???}

%Please use \dag to cite the ESI in the main text of the article.
%If you article does not have ESI please remove the the \dag symbol from the title and the footnotetext below.

%%%END OF FOOTNOTES%%%

%%%MAIN TEXT%%%%
\section{Introduction}
%What is active matter%

Group formation and collective motion in form of swarms or flocks is a hallmark of living systems on length scales from bacteria and sperm to school of fish, flocks of birds, and animal herds. \cite{ball:08,rama:10,laug:09,elge:15,shae:20,gomp:20} This behavior often emerges without central control and is rather governed by the response of individuals to the action of other group members or agents. Arising patters and structures not only depend on the physical interactions between the various agents of an ensemble, but are often governed by a nonreciprocal information input, e.g., by visual perception in case of birds and animals, processing of this information, and active response. Unravelling the underlying mechanisms and principles not only sheds light onto the behavior of biological systems, but provides concepts to design functional synthetic active systems, which are able to adopt to environmental conditions and perform complex tasks autonomously. \cite{Alan2020}

In dry active matter systems --- absence of hydrodynamic interactions --- self-propulsion and volume-exclusion lead to motility-induced phase separation (MIPS), as demonstrated in simulations of systems of active Brownian particles (ABPs) \cite{fily:12,bial:12,redn:13,wyso:14,cate:15,elge:15,marc:16.1,bech:16,digr:18} and in experiments. \cite{theu:12,thut:11,pala:13,butt:13,bechinger2016active} The presence of an embedding fluid, implying long-range hydrodynamic interactions, can suppress MIPS, \cite{mata:14,thee:18} but can also give rise to large scale collective effects\cite{ishi:08,alar:13,kyoy:15,alar:17,yosh:17} --- which depend on the presence or absence of thermal fluctuations\cite{thee:18} --- and swarming motility. \cite{qi:22}

Vision is the primary sensor for birds, and, as is established by now, a bird in a flock responds mainly to positions and motion of its seven nearest neighbors. \cite{ball:08.1,cava:14,ball:08,pear:14}  The Vicsek model for flocking accounts for such  interactions in a coarse-grained manner by a velocity alignment mechanism between nearby neighbors.\cite{vics:95} By varying the mean density, the strength of the reorientational noise, and the interaction radius, a wide spectrum of structures is obtained, such as global polar order or flocks.\cite{gine:10,gine:16,chat:20}   
This suffices to build up long-range orientational order in two dimensions \cite{tone:95,tone:98} and giant density fluctuations. \cite{rama:10} 

Alternative approaches for emergent collective behaviors without velocity alignment have been proposed, relaying on repulsion-attraction interactions, \cite{roma:09} elasticity-based interactions, \cite{ferr:13}, visual interactions with \cite{pear:14} and without coalignment, \cite{barb:16,bast:20} and aggregation based on chemical gradients. \cite{soto:14.1}  Not all models yield necessarily flocking, but particular phases and collective motion emerge, which are absent in ``dumb'' ABP-type systems, such as ''vision''-based aggregation and cluster formation, and worm-like structures as in Ref.~[\onlinecite{barb:16}].     

In this article, we study structure formation in systems of self-steering particles by applying the minimal cognitive flocking model proposed in Ref.~[\onlinecite{barb:16}]. Here, ABPs are additionally equipped with vision-based sensing, which allows them to detect the instantaneous positions of neighboring ABPs within a vision cone (VC) and a short-distances cutoff radius. The relaxation of their propulsion directions is governed by a redirection force of limited magnitude, implying a limited maneuverability, toward the centre of mass of the detected particles, which is in competition with rotational noise. Since the particles are able to sense and respond to their neighbors, we denote them as intelligent active Brownian particles (iABPs). The iABPs possess no memory and  lack the velocity alignment of the Vicsek model. As an extension to previous studies, \cite{barb:16} we take into account excluded-volume interactions between the iABPs and solve the underdamped equations of motion in the presence of thermal noise for their translational motion. 

Our simulations reveal various additional phases compared to those already presented in Ref.~[\onlinecite{barb:16}], which originate from excluded-volume interactions, such as hexagonally close-packed aggregates, fluid-like aggregates, worms, worm-aggregate coexistence, and dilute and diffuse clusters at different parameters sets. The strength of the response to the visual signal, the vision angle, packing fraction, cutoff range, and active propulsion determine the location and extent of these phases in the phase diagram.

Moreover, our analysis of the cluster-growth process from a dilute isotropic state via nucleation and merging of smaller clusters into larger ones exhibits distinct difference to the process for systems of short-range attractive passive colloids.\cite{evans1999colloidal} Specifically, the critical exponent of the dependence of the characteristic time for cluster growth on the number of particles in the cluster is larger than that of the passive colloidal system. The implemented perception rule slows down clustering and cohesion, which is attributed to the nonreciprocal visual interactions and the resulting directed motion.

\section{Model and simulation approach }

We consider a two-dimensional systems of  $N$ perceptive and responsive ``intelligent'' active Brownian particles (iABPs) at positions $\bm r_i(t)$ ($i=1,\ldots, N$) at time $t$, which are propelled by the active force $\bm F^{a}_i(t) = \gamma v_0 \bm e_i(t)$ with the velocity $v_0$ along the direction $\bm e_i(t)$. Their equations of motion are given by \cite{das2018confined}
\begin{equation} \label{equation:1}
m \ddot{{\bm r}_i} = -\gamma \dot{{\bm r}_i}   + \gamma v_0 \bm{e}_i + \bm F_i +  \bm{\varGamma}_i(t),
\end{equation}
with the mass $m$, the friction coefficient $\gamma$, and the transitional Gaussian white noise $\bm \varGamma_i(t)$ of zero mean and the second moments $ \langle \varGamma_{\alpha i}(t)\varGamma_{\beta j}(t') \rangle = 2 \gamma k_B T  \delta_{ij} \delta_{\alpha \beta} \delta(t-t')$  ($\alpha, \beta \in \{x,y\}\}$). The forces $\bm F_i$ account for excluded-volume interactions between the iABPs in terms of the short-range, truncated, and shifted Lennard-Jones potential
\begin{align}
     U(r)= \begin{cases}  \displaystyle  4 \epsilon \left( \left(\frac{\sigma}{r} \right)^{12} -\left(\frac{\sigma}{r} \right)^{6}\right) + \epsilon, & r\leq 2^{1/6} \sigma \\ \mbox{0,} & \mbox{otherwise} \end{cases}  ,
\end{align}
where $r = |\bm r|$ is the distance between two particles, $\sigma$ represents their diameter, and $\epsilon$ is the energy determining the strength of repulsion. Individual perception and noise lead to a change of the propulsion direction of the iABPs. Representing the propulsion directions in terms of polar coordinates,  $\bm e_i = (\cos \varphi_i , \sin \varphi_i)^T$, yields the equations of motion for the angles $\varphi_i$
\begin{equation} \label{equation:2}
  \dot{\varphi_i}= \frac{\Omega}{N_{c,i}} \sum_{j\in VC}e^{-r_{ij}/R_0} \sin({\phi_{ij}-\varphi_i})  +  \Lambda_i(t) ,
\end{equation}
where $N_{c,i}$ is the number of iABPS in the vision cone (VC), 
\begin{align} \label{eq:eq_2_norm}
  N_{c,i} =  \sum_{j\in VC}e^{-r_{ij}/R_0} .
\end{align}
 The  $\Lambda_i$ are Gaussian and Markovian stochastic processes  with zero mean,  the second moments  $\langle \Lambda_i(t) \Lambda_j (t') \rangle = 2D_R \delta_{ij} \delta(t-t')  $, and the rotational diffusion coefficient $D_R$. The sum in Eq.~\eqref{equation:2}  describes the preference of an iABP to move toward the iABPs in its ``vision'' cone (VC), with $\Omega$ the maneuverability strength. In a vision-based picture, the sum corresponds to the projection of the positions of all $N_c$ particles within the VC  onto the ``retina'' of particle $i$,  with $\phi_{ij}$ the polar angle of the vector $(\bm r_j - \bm r_i)/|\bm r_j - \bm r_i| = (\cos \phi_{ij}, \sin \phi_{ij})^T$ between particle $i$ and $j$. \cite{barb:16,pear:14} The condition for particles $j$ to lie within the vision cone of particle $i$ is 
 \begin{equation}
      \frac{\bm r_i- \bm r_j}{|\bm r_i- \bm r_j |} \cdot \bm e_i  \geq \cos(\theta) ,
 \end{equation}
 where $\theta$ --- denoted as vision angle in the following --- is the opening angle of the vision cone and $\bm e_i$ its orientation. \cite{barb:16} The exponential function  limits the range of the interaction with the characteristic length $R_0$. In addition, we limit the vision range to  $|\bm r_i - \bm r_j| \leq 4 R_0$ and treat all further apart particles as invisible.  
 
  \begin{figure}
 \subfloat{\includegraphics[width =\columnwidth]{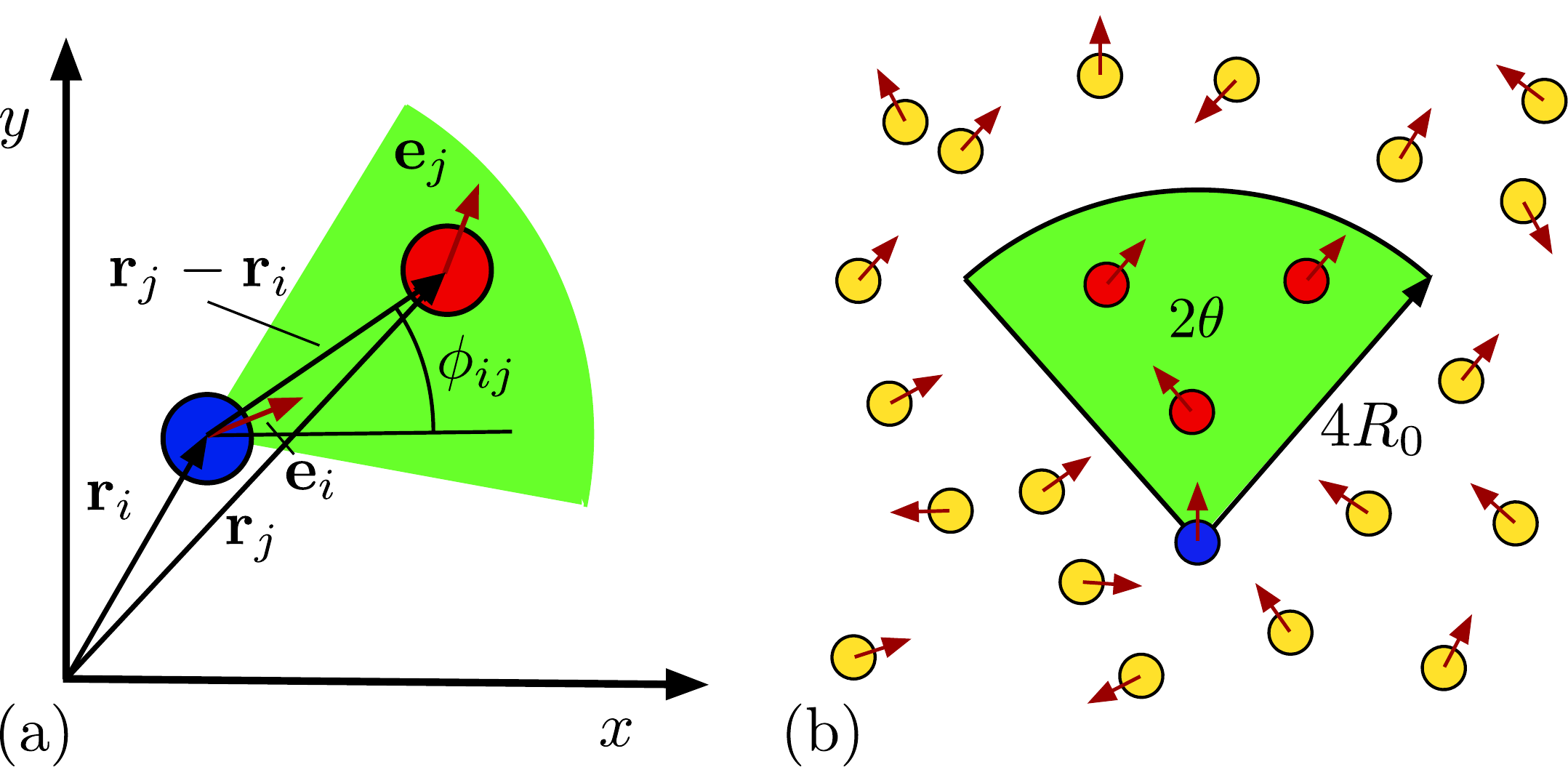}}
 \caption{(a) Graphical representation of the orientations ${\bf e}_i$ and ${\bf e}_j$ of the particles at  the positions ${\bf r}_i$ and ${\bf r}_j$, and the definition of the angle $\phi_{ij}$ of the connecting vector with the $x$ axis. b) Schematics showing the vision cone  of the blue particle with  the vision angle $\theta$ and the cutoff radius $4 R_0$. The vision cone is colored in green and the blue particle interacts with particles (red) within this cone. }
 \label{visual Representation}
 \end{figure}
%%%%%%%%%Figure 0 to be modified %%%%%%%
%%%%%%I will add figure here %%%%%%%
 
 \section{Parameters}

In the simulation, we measure energies in units of the thermal energy $k_BT$, lengths in units of $\sigma$, and time in units of $\tau = \sqrt{m\sigma^2/(k_BT)}$. The activity of the iABPs is characterized by the P\'eclet number
 \begin{equation}
     Pe=\frac{\sigma v_0}{D_T} , 
 \end{equation} 
 where $D_T = k_B T/\gamma$ is the translational diffusion coefficient. Explicitly, we choose $\gamma = 10^2 \sqrt{m k_BT/\sigma^2}$ and the rotational diffusion coefficient $D_R= 8 \times 10^{-2}/ \tau$, which yields the relation $D_T/(\sigma^2 D_R) = 1/8$. We set $\epsilon/k_BT=(1+Pe)$ to ensure an nearly constant iABP overlapping even at high activities. The iABP density is measured in terms of the global packing fraction $\Phi = \pi \sigma^2 N/(4L^2)$, with $L$ the length of the quadratic simulation cell. Periodic boundary conditions are applied and the equations of motion \eqref{equation:1} are solved with a Velocity-Verlet-type  algorithm suitable for stochastic system, \cite{gronbech2013simple}  with the time step $\Delta t = 10^{-3} \tau$. We perform $10^6$ equilibration time steps and collect data for additional $10^7$ steps. For certain averages, up to 10 independent realizations are considered.
 
 The choice of the large friction coefficient ensures that inertia does not affect the structural and dynamical properties as long as there are no dense clusters are formed. In case of emerging dense clusters, the position of phase boundaries weakly depends on the presence of inertia.      
 
 If not indicated otherwise, the number of particles is $N=625$, the characteristic radius $R_0 = 1.5\sigma$, $\Omega =  5/\tau$, and the opening of the vision cone $\pi/24 \leq \theta \leq \pi/2$. 

Initially, the iABPs are typically arranged on a square lattice, with iABPs distances equal to their diameter $\sigma$ in the center of the periodic simulation square.

\section{Structural and dynamical properties}

\subsection{Low packing fraction}

\subsubsection{Phases and phase diagrams} \label{sec:phases_low}

\begin{figure*}
\includegraphics[width = \textwidth]{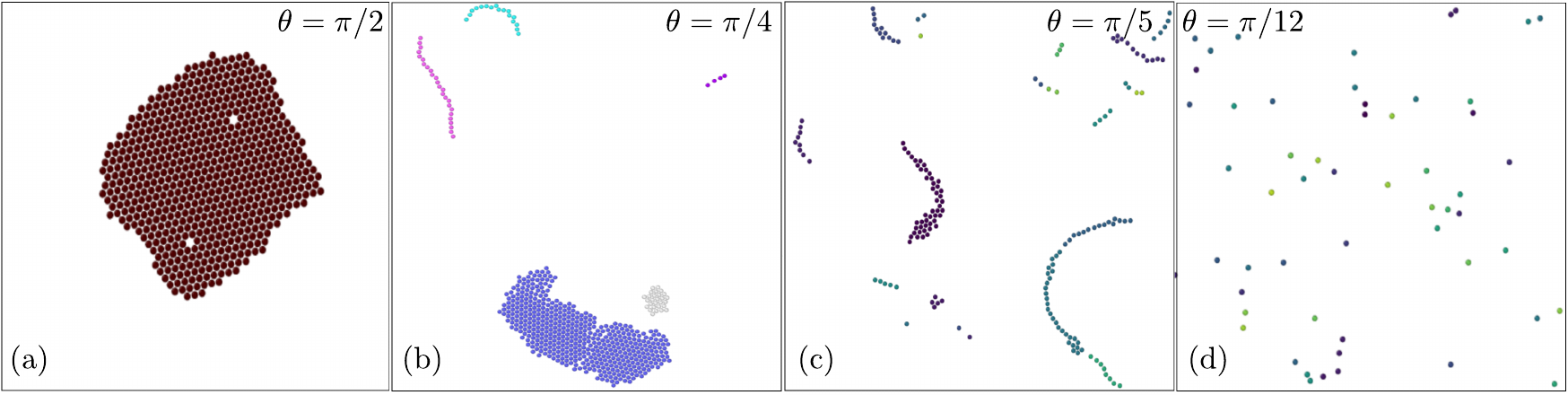} 
\caption{Snapshot of iABP structures for various vision angles: (a) hexagonal-closed packed structure, $\theta =\pi/2$, (b) worm-aggregate coexistence, $\theta = \pi/4$, (c) worms, $\theta = \pi/5$, and (d) dilute fluid, $\theta = \pi/12$.  The P\'eclet number is  $Pe=200$, the strength of the perceptive interaction $\Omega =5$, and the packing fraction $\Phi= 0.00785$.  The various colors represent different clusters or worms. See movies M1 and M2 (ESI$\dag$).}
\label{Screenshot1}
\end{figure*}

Figure \ref{Screenshot1} displays snapshots of emerging structures for various vision angles at the very small packing fraction $\Phi = 0.00785$, indicating  an aggregates, mobile worms, worm-aggregate coexistence, and a dilute phase. 

To characterize the various phases, we calculate the cluster-size distribution function  \cite{Mario2018,allen2017computer}
\begin{equation}\label{eq:cluster}
  {\cal N} (n)= \frac{1}{N_N} np(n) ,
\end{equation}
which is the average fraction of particles in clusters of size  $n$, and $p(n)$ is the number of a clusters of size $n$. 
The distribution is normalized such that $\sum_n {\cal N}(n) = 1$, which determines $N_N$.
We use a distance criterion to define a cluster, where an iABP belongs to a cluster when its distance to another iABP of the cluster is within a radius $\sigma_0$. Since we mainly focus here on dilute systems and worms, we set $\sigma_0 = 2  \sigma$. 

\begin{figure}[t]
    \includegraphics[width=.48\textwidth]{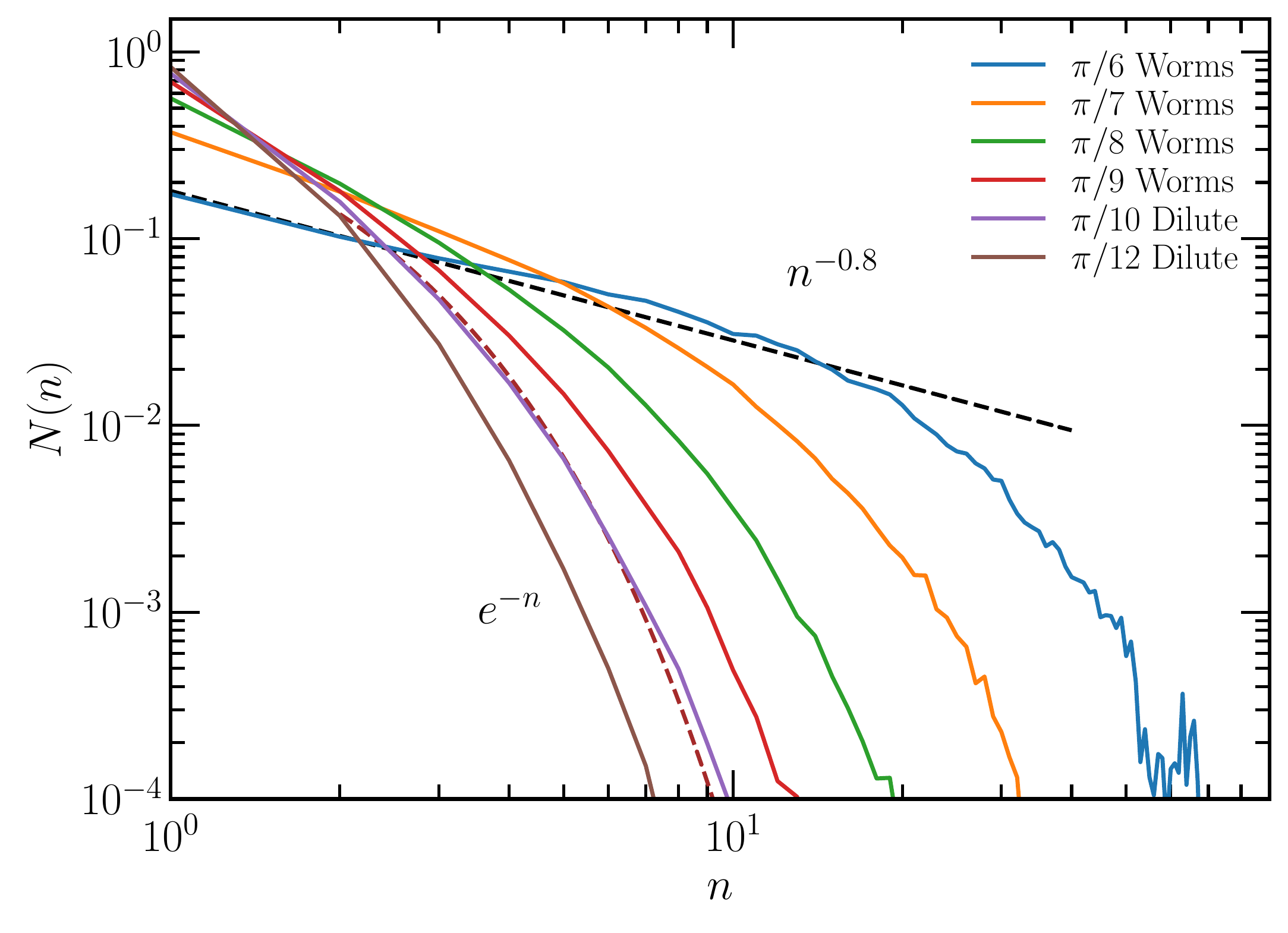}
    \caption{Cluster-size distribution function ${\cal N} (n)$ as a function of the cluster size $n$ at $Pe=200$,  $\Omega=5 $, and the packing fraction $\Phi = 7.85\times 10^{-3}$. The black dashed line indicates a power-law decay,  and the brown dashed line the exponential decay $e^{-n/n_0}$, with $n_0=1$. }
        \label{Fig:Cluster}
\end{figure}

Figure \ref{Fig:Cluster} presents the cluster-size distribution function for various vision angles. As long as the vision angle $\theta \lesssim \pi/10$,  ${\cal N}$ decays exponentially with a characteristic cluster size $n_0$. For  $\theta > \pi/10$, a power-law decay at small $n$ appears, followed by an exponential cut-off for larger clusters. This indicates longer-range correlations in the system, but no large clusters. \cite{Mario2018,qi:22} For the current system, the power-law is most pronounced at $\theta = \pi/6$, where $ {\cal N}(n) \sim n^{-\mu}$, with $\mu=0.8$ as long as $n \lesssim 20$. 

\begin{figure}[t]
 \includegraphics[width=.48\textwidth]{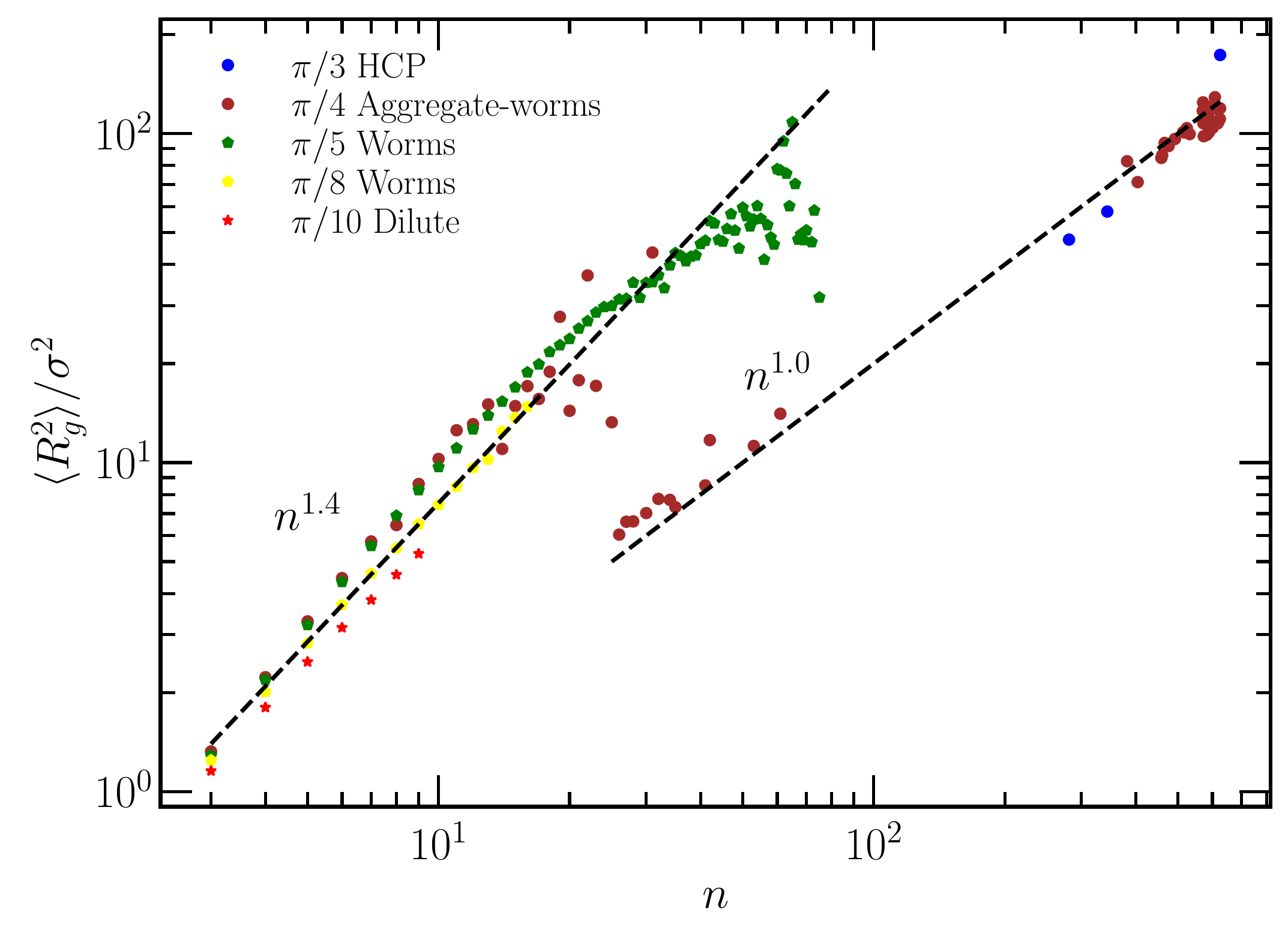}
 \caption{Radius of gyration as a function of the cluster size at $Pe=200$,  $\Omega=5$, and the packing fraction $7.85\times 10^{-3}$ for various vision angels. The two distinct  power-law regimes characterize worms ($n^{1.4}$) and aggregates ($n$).  Worm-aggregate coexistence is  indicated by a crossover from the worm power-law at smaller clusters to the aggregate one at larger clusters. }
 \label{Fig:Radius}
 \end{figure}
 
To classify the various phases of Fig.~\ref{Screenshot1}, we calculate the radius of gyration, 
\begin{equation}
  \langle R_g^2 \rangle  = \frac{1}{n} \sum_{i=1}^{n} \left(\bm r_i - \bm r_{cm} \right)^2  ,
\end{equation} 
of the cluster of size $n$, where $\bm r_{cm}$ is the center-of-mass position of the cluster. In order to avoid counting configuration which occur very rarely, we only consider realizations which appear in more than $1\%$ of recorded  configurations.  As displayed in Fig.~\ref{Fig:Radius}, the data collapse on two branches with the distinct power laws $\langle R_g^2 \rangle \sim n^{1.4}$ and $n^1$, respectively, for the various vision angles. All data points for $\theta  \lesssim \pi/5$ are approximately on the line with the exponent $n^{1.4}$, whereas  the radii of gyration for $\theta = \pi/3$ exhibit a linear increase with $n$. The system for $\theta = \pi/4$ shows different features for $n \lesssim 20$ and $n \gtrsim 20$.  
 
For a disc-like dense packing of $n$ iABPs, its radius, $r$, increases as $r^2 \sim n$. Hence, the linear increase of $\langle R_g^2 \rangle$  is consistent with a dense aggregate of iABPs. Thus, we expect dense clusters of iABPs for $\theta = \pi/3$ and $\pi/4$ for larger $n$, which is confirmed by visual inspection (Fig.~\ref{Screenshot1}(a)). Larger exponents correspond to more open structures. In particular, a linear arrangement of iABPs in a rod-like manner  would correspond to $\langle R_g^2 \rangle \sim n^2$. Visual inspection indeed shows the presence of wormlike structures, which, however, are not rodlike, but rather curvilinear, hence, $\langle R^2_g \rangle \sim n^{\alpha}$ with $\alpha <2$ (Fig.~\ref{Screenshot1} (c)). In the system with $\theta = \pi/4$, wormlike structures and dense aggregates coexist (Fig.~\ref{Screenshot1}(b)). 

\begin{figure*}
 \includegraphics[width = \textwidth]{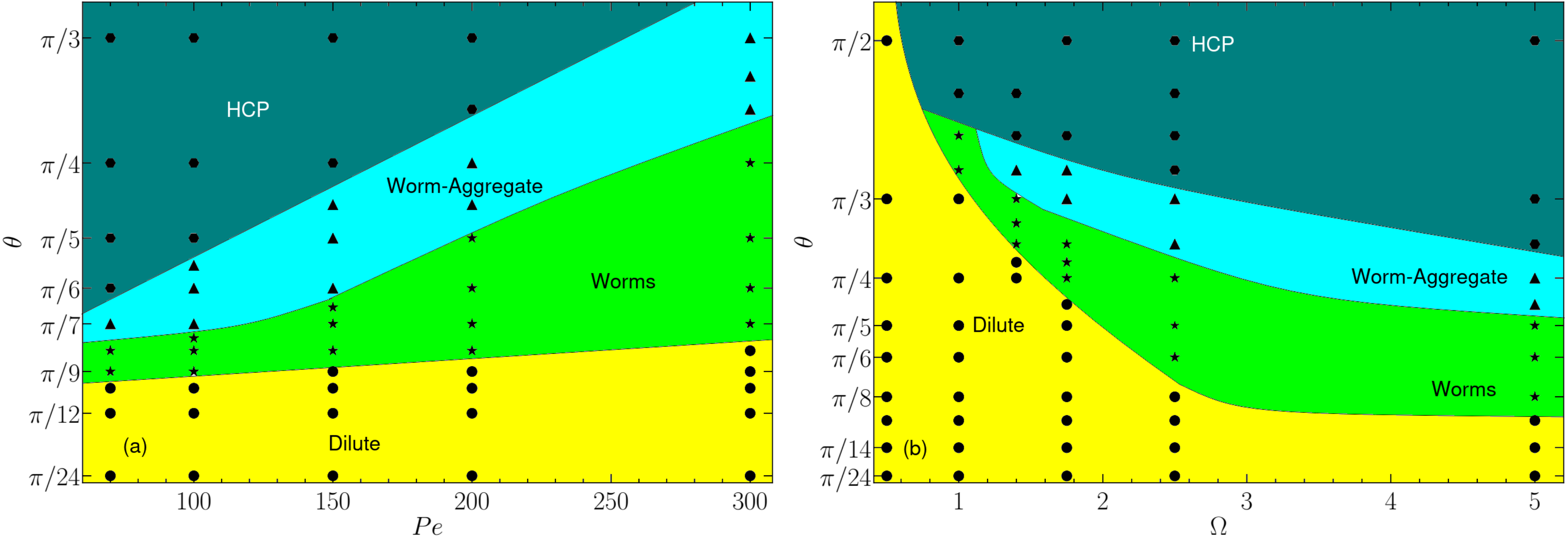} 
 \caption{(a) Vision angle-P\'eclet number ($\theta -Pe$) state diagram for the packing fraction $\Phi= 0.00785$,  $\Omega=5$, and $R_0=1.5\sigma$.  (b) Vision angle-$\Omega$ ($\Omega-Pe$) state diagram  for $Pe=200$, and otherwise the same parameters as in (a). The individual phases are indicated by different colors and symbols.  HCP: navy ${\hexagofill}$,  worm-aggregate: blue ${\blacktriangle}$,  worm: green {$\bigstar$}, dilute: yellow ${\CIRCLE}$. The ``phase'' boarders are guides for the eye.}
 \label{Phase_diagrams_low_density}
 \end{figure*}

Based on the radius-of-gyration results and the cluster-size distributions, we construct the phase diagram of Fig.~\ref{Phase_diagrams_low_density} for various $Pe$ and $\Omega$.  Four phases can be clearly identified: (i) a dilute phase  ($\theta \lesssim \pi/10$ in Fig.~\ref{Phase_diagrams_low_density}(a)),  (ii) a phase of mobile worms for a narrow range of somewhat larger $\theta$, (iii) a coexistence of worms and aggregates, denoted as worm-aggregate phase,  and (iv) a phase with hexagonally closed-packed aggregates, denoted as HCP phase, at large $\theta$. 

In the dilute phase,  $\theta \lesssim \pi/10$, the system is homogeneous and isotropic. The vision angle and range are too small to allow for persistent sensing of neighboring particles, and the noise dominates over maneuverability. For the parameters of Fig.~\ref{Phase_diagrams_low_density}(a), this behavior depends only weakly on $Pe$, with an extension of the dilute phase toward somewhat larger $\theta$ at larger P\'eclet numbers. As $\theta$ increases, worms appear, where an iABP is trailed by several others. Similar structures have previously been observed in Ref.~[\onlinecite{barb:16}] for pointlike particles.  As indicated in Figs.~\ref{Screenshot1} and \ref{Fig:Radius}, the worms can be quite long, comprising up to $80$ iABPs. Here, the truncated perception cone is sufficiently large to allow for a persistent tracking and chain formation. The directional dynamics of the worm is determined by the leading iABP, since it does not see other iABP and the orientational dynamics is governed by its rotational noise. The stretch of the worm phase depends on the P\'eclet number and extents to larger $\theta$ at higher $Pe$. Even wider vision cones lead to formation of aggregates, which coexist with worms --- the worm-aggregate phase. The presence of a wider vision cone implies a preferred motion of more particles toward each other, and the emergent locally higher density stabilizes the clusters.  Again, the extent of this phase depends on $Pe$. Finally, a densely packed aggregate is obtained for large $\theta$, comprising essentially all iABPs. Our analysis shows that the local structure in the aggregate is hexagonally closed packed (Fig.S4 ESI$\dag$). iABPs always discern sufficiently many neighbors to stay close to them. Only for large $Pe$, iABPs are able to escape, trailing other particles, and form worms.  

Here, a comment on the stationary-state properties of the observed structures is in order. The identified worm-aggregate, worm, and dilute phase are stationary, and do not depend on the initial arrangement of the ABPs. These structures are very dynamic, they form, break apart, and reform. The situation is less clear for the HCP phase. Here, the initial arrangement never breaks apart, and it is not clear whether it is stable or not. However, when starting from a dilute-phase as initial condition, clusters are formed, which slowly merge into larger ones as described in more detail in Sec.~\ref{sec:cluster_growth}, but the possible final state of a single large cluster cannot be observed in simulations in general due to the very slow merging process.      

The  extension of the various phases in the phase diagram depends on the maneuverability strength $\Omega$, as displayed in    
Fig. \ref{Phase_diagrams_low_density}(b). At very small value of $\Omega$, noise dominates the orientational dynamics, and the iABPs are not able to persistently move toward sensed other iABPs --- the system is in the dilute phase. For larger $\Omega$, the maneuverability begins to dominate over rotational noise, iABPs can move into the sensed direction, and, depending upon the vision angle, can form HCP,  worm-aggregate, worm phases. 

%%Figure for the phase diagram%%%%%%%%%%%%%%
 %%%%%%%%%%%%%%%%%%%%%%%%%%%%%%%%%%%%%%%%%%%%%%%%%%%%%%%%%%%%%%%%%%%%%%%%%%%%%%%%%%%%%%%%%%%%%%%%
 %%%%%%%%%%%%%%%%%%%%%%%%%%%%%%%%%%%%%%%%%%%%%%%%%%%%%%%%%%%%%%%%%%%%%%%%%%%%%%%%%%%%%%%%%%%%%%%%%%%%%%

\subsubsection{Collective dynamics}

\begin{figure}
    \includegraphics[width=.48\textwidth]{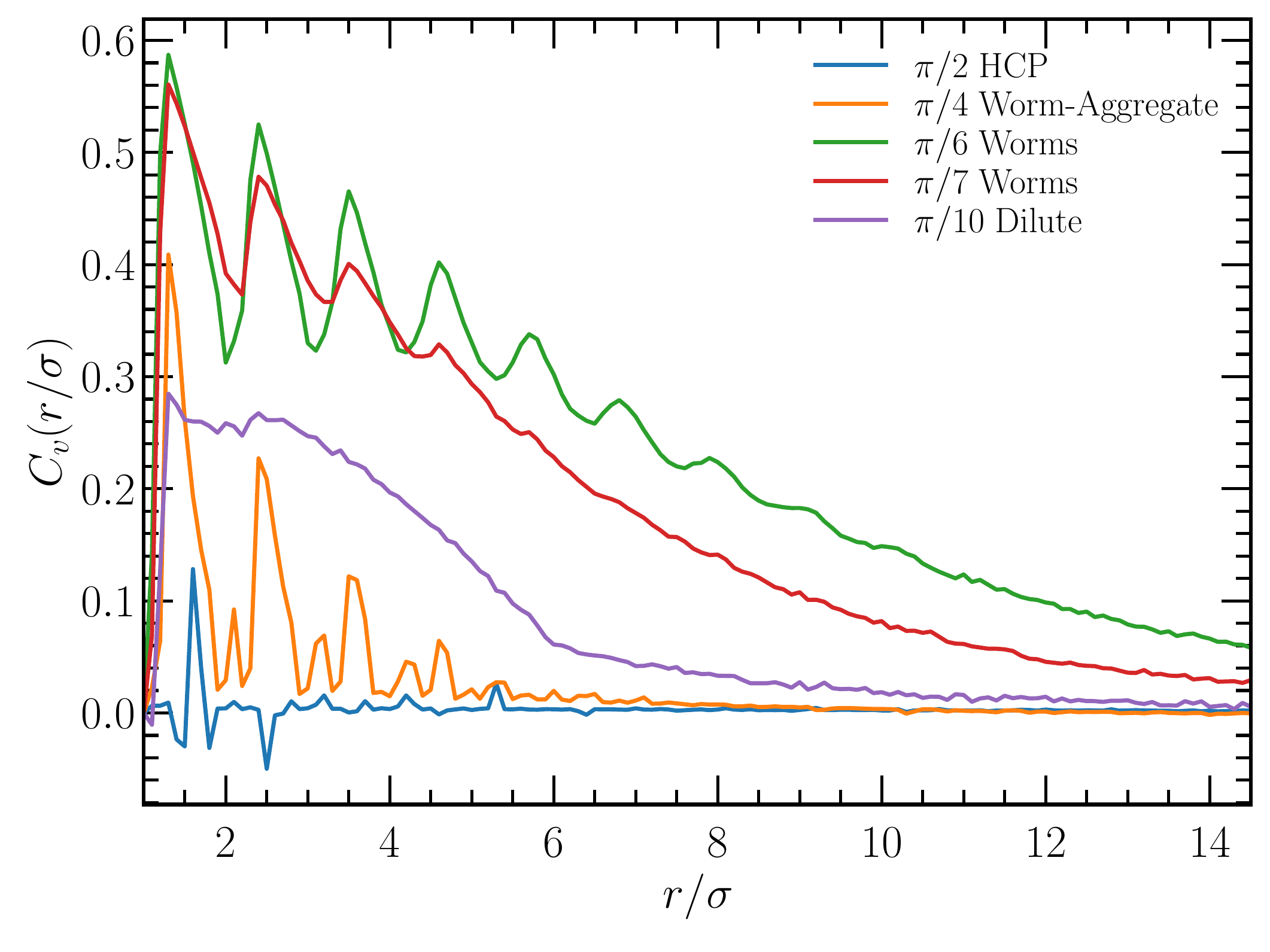}
    \caption{Spatial velocity autocorrelation function of ABPs at $Pe=200$,   $\Omega=5$, vision range $R_0=1.5\sigma$, and the packing fraction $\Phi =7.85 \times 10^{-3}$.}
    \label{Fig:Velocity_Spatial}
\end{figure}

We characterize dynamical correlations by the spatial velocity correlation function  \cite{wysocki2014cooperative}
\begin{align} \displaystyle 
     C_{v}(\bm r) = \frac{\displaystyle  \sum_{i,j\ne i} \left \langle \bm v_i  \cdot \bm v_j \delta(\bm r-|\bm r_i-\bm r_j|) \right \rangle}
    {\displaystyle c_0  \sum_{i,j\ne i} \left \langle \delta(\bm r-|\bm r_i-\bm r_j|) \right\rangle} ,
\end{align}  
where $\bm v_i$  is the velocity of particle $i$ and $c_0=\langle \bm \sum_i v_i^2 \rangle/ N$.
Figure~\ref{Fig:Velocity_Spatial} presents $C_v(\bm r)$ for various vision angles. At the angle $\theta=\pi/2$, in the HCP phase,  there is no significant spatial velocity correlation. The propulsion directions are independent, because iABPs inside a cluster see the same environment in any direction. The finite value of $C_v$ for $\theta =\pi/10$, and its extent over several iABP diameters, indicates significant velocity correlations already in the dilute phase. Hence, the dilute phase is different from the fluid phase of a passive system, as is also reflected in the spatial pair distribution function (Fig. S4). Long-range correlations are inherent in worms. Naturally, the trailing of iABPs in worms is only possible with strong velocity correlations. The maxima correspond to preferred distances between iAPBs at multiples of $1.25 \sigma$, as also displayed by the pair distribution function. Minima appear approximately at multiples of $\sigma$. The distance $1.25 \sigma$ is consistent with visual inspection, and suggest that the distance between subsequent iABPs in a worm is larger than their diameter, i.e., the typically do not touch.  Worms are often nonlinear  assemblies, with touching particles forming clusters moving together. The velocities of such additional iABPs are not necessarily fully aligned along the main worm direction, giving raise to smaller values of the velocity correlation function. This explains the oscillations of $C_v$ for $\theta =\pi/6$ and $\pi/7$. 

\subsubsection{Mean square displacement} \label{sec:msd_low}

The translational motion of the ABPs is characterized by their mean-square displacement (MSD)
\begin{equation}\label{MSD}
    \langle \bm r^{2}(t) \rangle= \frac{1}{N} \sum_{i=1}^N \left\langle \left( \bm r_i(t)- \bm r_i(0) \right)^2 \right\rangle .
\end{equation}
Theoretical calculations for individual overdamped ABPs in two dimensions yield  \cite{hows:07,elgeti2015physics}
\begin{equation}\label{MSD_2}
\langle \bm r^{2}(t) \rangle =4D_T t + \frac{2v_0^2}{D_R^2} \left( D_R t -1 + e^{-D_R t}\right) ,
\end{equation}
with a short time diffusive regime, an active ballistic regime, and a crossover to an activity dominated diffusion for $D_Rt>1$. 
\begin{figure}
\includegraphics[width=.48\textwidth]{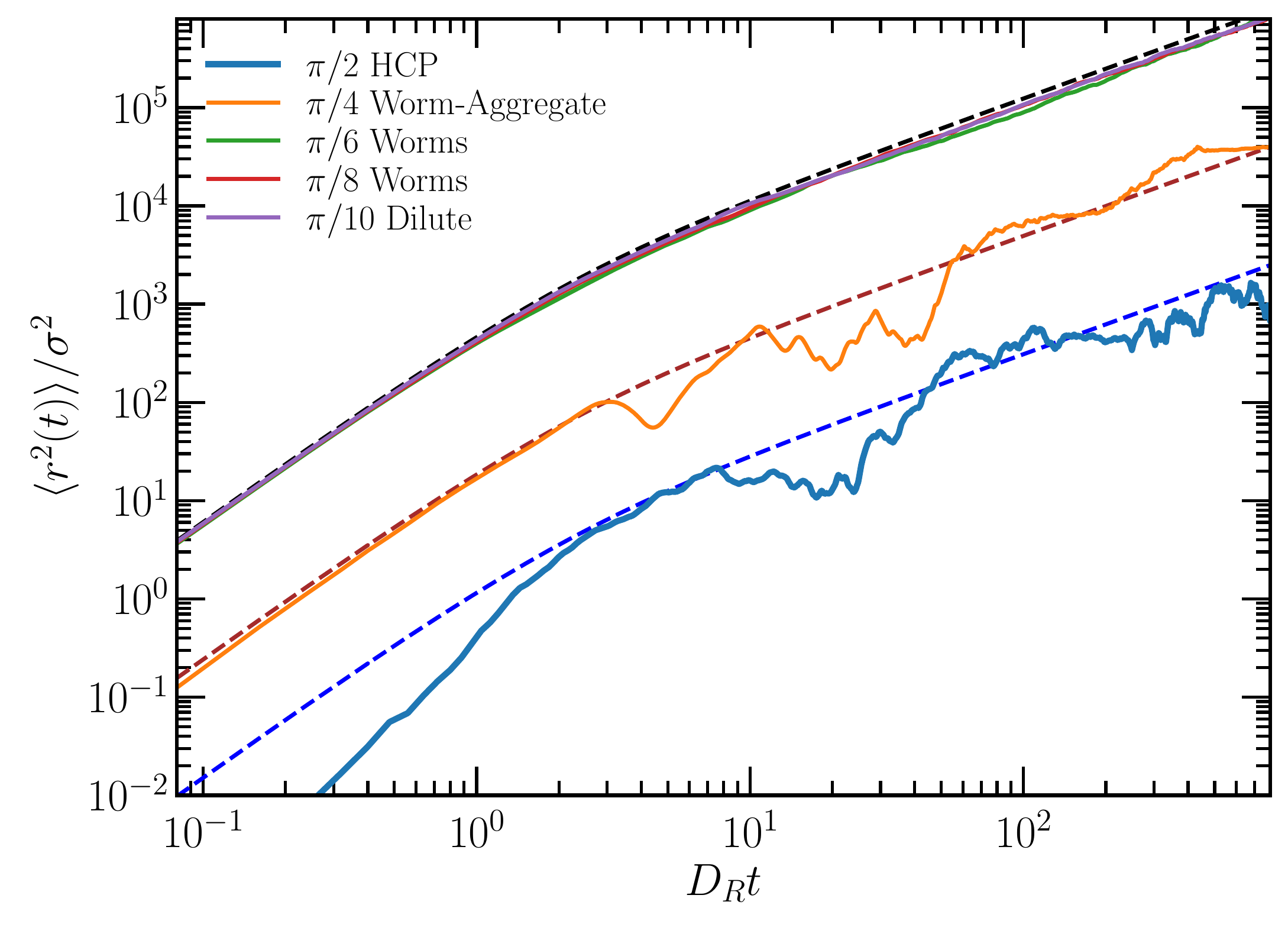}
\caption{Mean-square displacement of iABPs as a function of time at $Pe=200$,   $\Omega=5$, vision range $R_0=1.5 \sigma$, and the packing fraction $\Phi = 7.85 \times 10^{-3}$. The dashed lines are fits of Eq.~\eqref{MSD_2}, without the passive diffusion term $ 4D_Tt$.  }
\label{fig:MSD}
\end{figure}
Figure~\ref{fig:MSD} displays MSDs for various vision angles, $Pe=200$, and $\Omega = 5$. As long as $\theta \lesssim  \pi/6$, which corresponds to the worm and dilute phase,  the MSDs are well described by the expression of independent and noninteracting iABPs, with the set values $D_R=0.08 \tau$ and $Pe=200$.  This is not surprising, because the leading ABP of a worm does not see any other ABP for most of the time, and thus, behaves as a free ABP.  The trailing ABPs follow the ``leader'' on essentially the same random trajectory. Thus, worm conformations roughly resemble the trajectories of individual ABPs.  The presence of aggregates changes the diffusive behavior. In the worm-aggregate phase, the MSD is also well described by the theoretical expression for ABPs, but with the reduced active velocity $v_0' =  0.4 \sqrt{k_BT/m} = v_0/5$, determined by the fraction of iABPs in aggregates.  The dynamics of the ABPs in the aggregates are slower than free ABPs, but  ABPs are able to split off, move like independent ABPs, and rejoin dynamical clusters. In average, we obtain an ABP dynamics with a reduced effective P\'eclet number. 

In the HPC phase for $\theta \leq \pi/2$, we still observe long-time active diffusion for $D_Rt>1$, with the active velocity $v_0' =  0.1 \sqrt{k_BT/m} = v_0/20$, i.e., a $20$ times smaller active P\'eclet number. Here, the aggregate moves as a whole. 
Calculations for $N$ bound active Brownian particles yield an active diffusion coefficient proportional to $1/N$,\cite{eise:16} which corresponds to an effective active velocity $v_0' \sim/\sqrt{N}$. With about $650$ particle in a cluster, the reduction by the factor $20$ of $v_0$ is consistent with the picture of a cluster moving actively as a whole.

\subsection{Higher packing fraction}

\subsubsection{Phases and phase diagrams}

\begin{figure*}
\includegraphics[width = \textwidth]{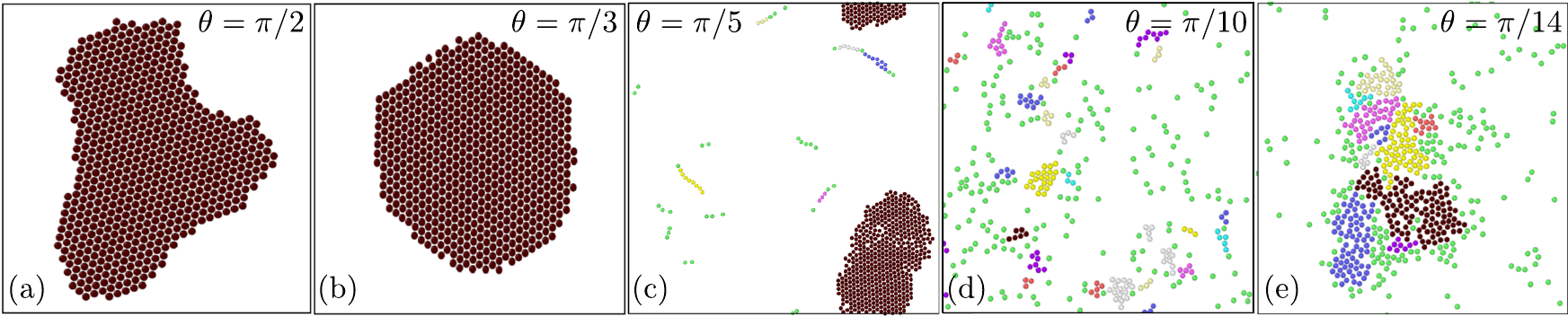}
\caption{Snapshot of iABP structures for various vision angles: (a) hexagonal-closed packed (HPC) structure, $\theta =\pi/2$, (b) hexagonal-closed packed (HPC) structure, $\theta =\pi/3$, (c) worm-aggregate coexistence, $\theta = \pi/5$, (d) dispersed cluster (DC) phase, $\theta = \pi/10$, for the P'eclet number $Pe=200$, and (e) aggregate phase , $\theta = \pi/12$, for $Pe=10$.  The strength of the perceptive interaction is $\Omega =5$, and the packing fraction $\Phi=0.0785$. The various colors represent different cluster. See movies M3, M4, and M5 (ESI$\dag$).}
\label{Screenshot2}
\end{figure*}

Figure \ref{Screenshot2} displays snapshots of iABP structures for various vision angles and the packing fraction $\Phi =0.0785$. We obtain similar phases as in Fig.~\ref{Screenshot1} with aggregates and worms (see also Fig.~S1 ESI$\dag$). In addition to Fig.~\ref{Screenshot1}, two more phase appear,  a dispersed cluster phase (DC) comprised of small clusters (Fig.~\ref{Screenshot2}(d)), and a fluid-like aggregate with rather mobile groups, which we denote as aggregate phase (Fig.~\ref{Screenshot2}(e)). 

\begin{figure}
    \includegraphics[width=.48\textwidth]{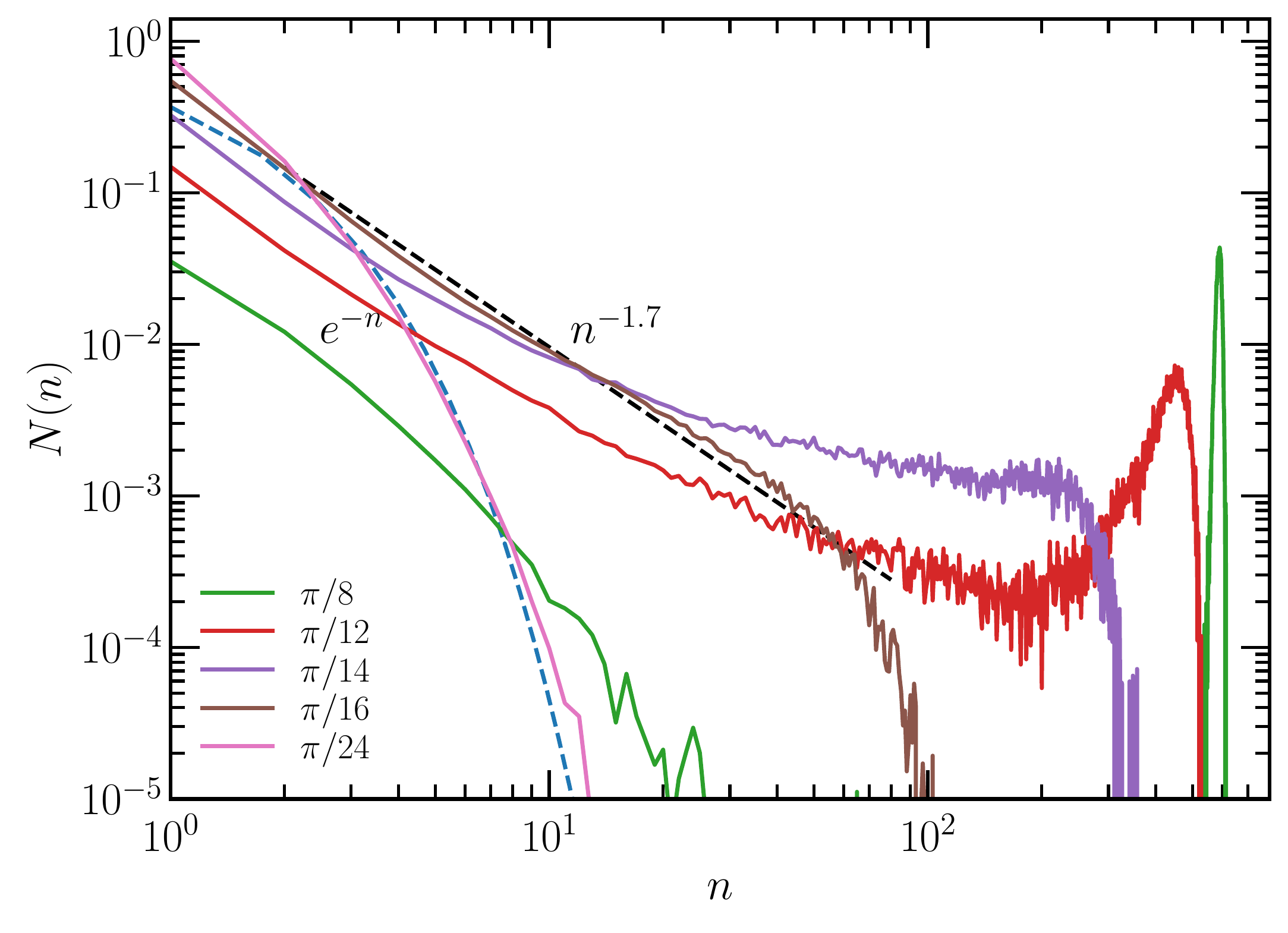}
    \caption{Cluster-size distribution function ${\cal N} (n)$ as a function of the cluster size $n$ at $Pe=10$,  $\Omega=5 $, and the packing fraction $\Phi = 7.85\times 10^{-2}$. The dashed line indicates a power-law decay. Notice that the radius to identify  ABPs belonging to a cluster  is $\sigma_0 = 1.2 \sigma$.}
   \label{Fig:Cluster_2}
\end{figure}

The cluster-size distribution function of Fig.~\ref{Fig:Cluster_2} shows the crossover from a fluid-like phase with an exponentially decaying ${\cal N}$ for $\theta < \pi/16$ to an bimodal distribution characteristic for a phase separated system. \cite{qi:22} The angle $\theta = \pi/16$ is close to the boundary separating the two phases. Note that we use  $\sigma_0=1.2 \sigma$ in this case to identify iABPs belonging to a particular cluster.  

\begin{figure}
\includegraphics[width=.48\textwidth]{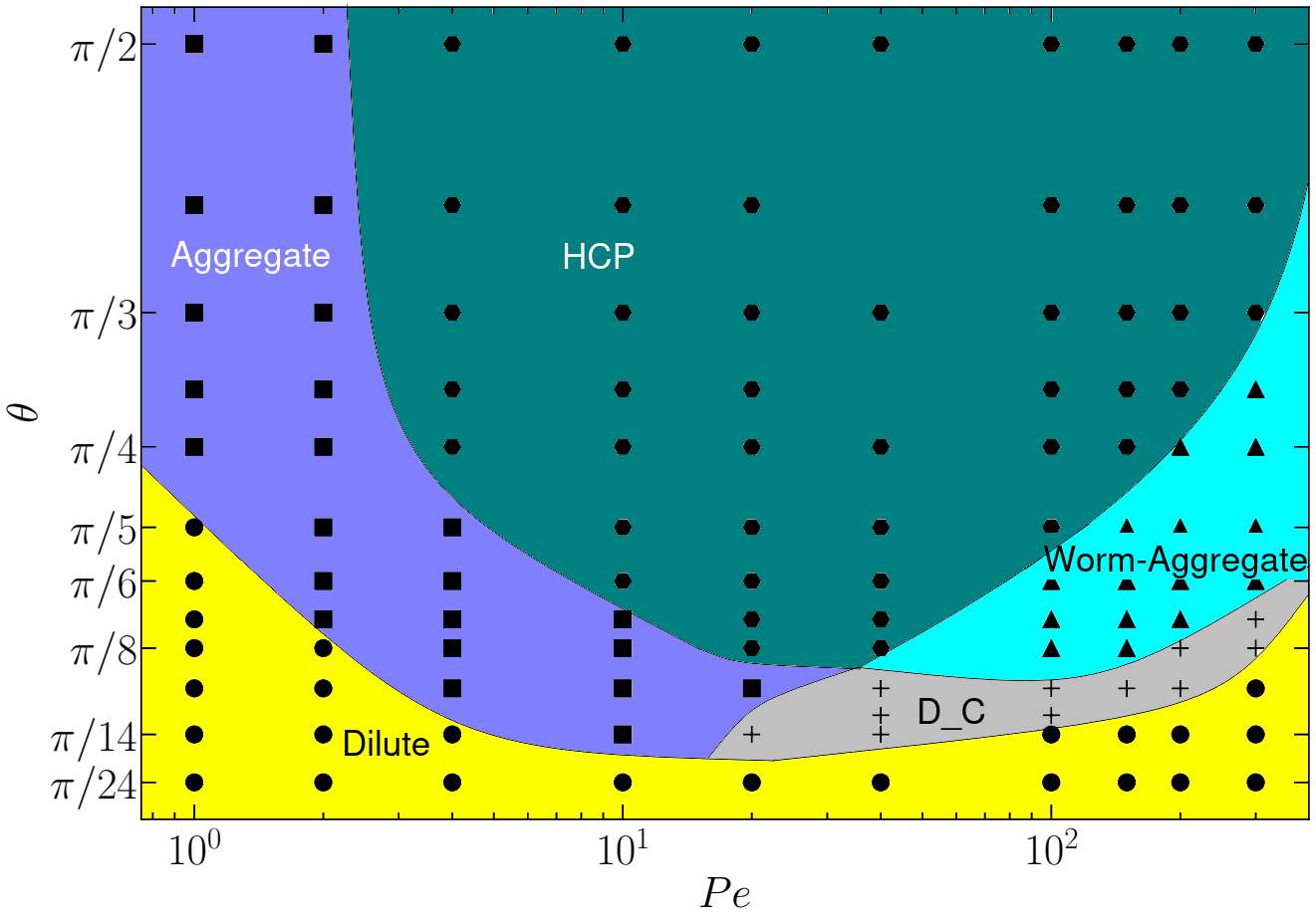}
\caption{Vision angle-P\'eclet number ($\theta -Pe$) state diagram for the packing fraction $\Phi= 7.85\times 10^{-2}$,  $\Omega=5$, and $R_0=1.5\sigma$.  The individual phases are indicated by different colors and symbols.  HCP: navy ${\hexagofill}$,  worm-aggregate: blue ${\blacktriangle}$,  worm: green {$\bigstar$}, dilute: yellow ${\CIRCLE}$, distributed cluster (DC): grey +, aggregate: purple $\blacksquare$. The ``phase'' boarders are guides for the eye rather than strict boundaries.}
\label{Fig:Phase_Pe_vs_theta_high_den}
\end{figure}

Based on the cluster-size distribution function, visual inspection, and the hexagonal order parameter $q_6$ (Fig. S3)  \cite{Mario2018,stei:83,bial:12} --- specifically for separating the aggregate and HCP phase ---, we obtain the phase diagram in Fig.~\ref{Fig:Phase_Pe_vs_theta_high_den}.  Again, we find a HCP, worm-aggregate, and a dilute phase, however no worm phase. Evidently, higher densities enhance the preference of cluster and aggregate formation, because more particles appear in the vision cone.  At smaller $Pe< 50$, in the aggregate phase, the iABPs form smaller cohesive clusters, which remain close to each other, but are highly dynamic by resolving and reforming while in contact with a dilute phase (Fig.~\ref{Screenshot2}(e)). At low $Pe$, this phase extends up to $\theta =\pi/2$. With increasing $Pe$, and for $\theta > \pi/8$, a HCP phase is present. The more persistent motion implies stronger cohesion and formation of a denser aggregate. Even larger $Pe$ lead to melting of the HPC aggregates and a worm-aggregate phase appears as long as $Pe \gtrsim 50$ and $\pi/9 <\theta < 3 \pi/2$ (Fig.~\ref{Screenshot2}(c)).  Compared to the worm-aggregate phase at the lower density, the worms are typically shorter in length and shorter lived (Sec.~\ref{sec:phases_low}). For smaller vision angle, the second new phase, the dispersed cluster phase, appears at the high P\'eclet numbers, with small and short-lived clusters (Fig.~\ref{Screenshot2}(d)). 

The various phases not only depend on $Pe$, but also on the vision range $R_0$ and the maneuverability  $\Omega$. Figure S2 (ESI$\dag$) presents the dependence on $R_0$ and $\Omega$ for $Pe=10$. The three phases, dilute, aggregate, and HCP, as in Fig.~\ref{Fig:Phase_Pe_vs_theta_high_den}, are present for the range of considered values $R_0$ and $\Omega$. With increasing vision range and maneuverability strength, the transition to the aggregate and HCP phase shifts to smaller vision angels. This is to be expected, because by a large $R_0$, more iABPs are present in the vision cone, enhancing the tendency of clustering, and thus, cohesion. Similarly, at a larger $\Omega$, orientational noise is less important compared the maneuverability, which also enhances cohesion.

\begin{figure}[t]
    \includegraphics[width=.48\textwidth]{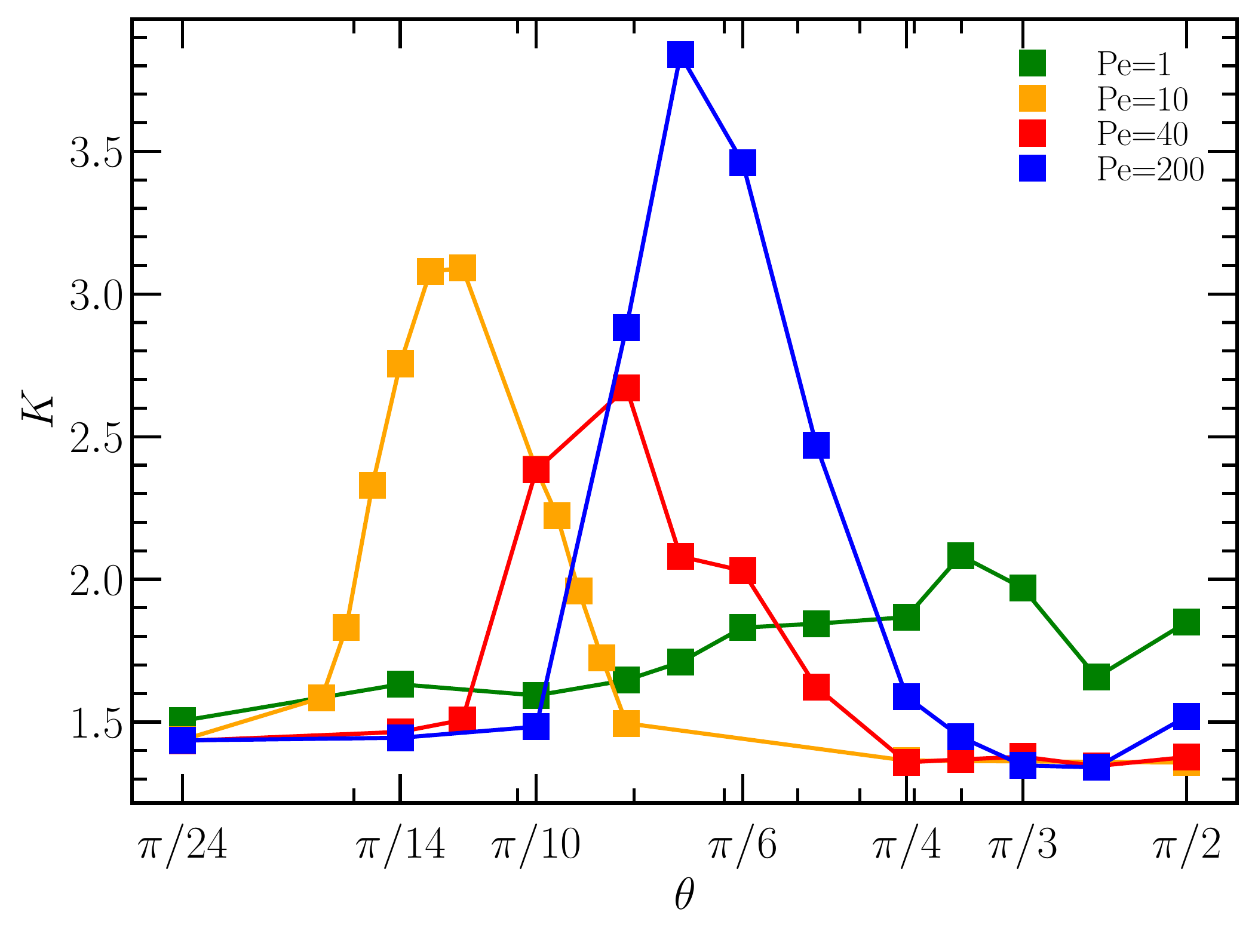}
    \caption{Kurtosis of the probability distribution of iABPs for various P\'eclet numbers, $\Omega=5$,  vision range $R_0=1.5 \sigma$, and the packing fraction  $\Phi=7.85\times 10^{-2}$ (symbols). The lines are guides for the eye.}
    \label{fig:non_guassian}
\end{figure}

\subsubsection{Shape of aggregates}

To characterize the shape of the iABP structures, we determine the kurtosis
\begin{equation}
    K = \frac{\langle \Delta \bm r^4 \rangle}{\langle \Delta \bm r^2 \rangle ^2} ,
\end{equation}
where 
\begin{align}
\Delta \bm r^{\zeta} = \frac{_{1}}{N} \sum_{i=1}^N \left(\bm r_i - \bm r_{cm} \right)^\zeta ,
\end{align}
with $\zeta = 2, 4$. The kurtosis of a Gaussian distribution is  $K =3$, that of  a uniform distribution within a square is $K=7/5$, and for a circle $K=4/3$. Respective larger values indicate deviations from homogeneity and isotropy, where larger variations imply larger $K$. 

Figure~\ref{fig:non_guassian} shows the kurtosis for various $Pe$. At very low vision angles, in the dilute phase, the ABPs are homogeneously distributed and $K\approx 1.4$ for all $Pe$, in agreement with the theoretical prediction $K=7/5$. For $Pe=1$, the $K$ are close to the isotropic value at small $\theta$, but increase with increasing $\theta$, indicating formation of inhomogeneities, consistent with the formation of rather mobile clusters.  

Similarly, for $Pe=10$ and $\theta \lesssim  \pi/20$ the system is isotropic and homogeneous. At $\theta \approx \pi/18$ , $K$ increases sharply and reaches the value $K \approx 3.2$ corresponding to a symmetric Gaussian distribution of iABPs (Fig.~\ref{Fig:Phase_Pe_vs_theta_high_den}). The value $K$ suggests that the shape of the aggregate is flexible and the iABPs are mobile, which agrees with the visual impression of a very dynamic, fluid-like cluster. Between $\theta = \pi/8$ and $\theta = \pi/2$, the phase transition from the aggregate phase to the HCP phase occurs, and $K$ approaches the value $K\approx 3/4$. Here, the iABPs are densely packed and uniformly distributed in a circular area much smaller than the simulation cell (Fig.~\ref{Screenshot2}(b)). 

The broad peak at $Pe=40$ for $\pi/14 < \theta < \pi/4$ comprises configurations in the distributed cluster, worm-aggregate, and HCP phase. The values $7/4 \lesssim K \lesssim 2.5$ point toward large variations in density or/and shapes of clusters. 

For $Pe=200$, $K$ increases as distributed clusters appear $(\theta \gtrsim \pi/10)$, reaches a maximum at $\theta \approx \pi/7$ in the worm-aggregate phase (Fig.\ref{Screenshot2}(c)). In the latter case, the structures are rather non-uniform, specifically the worms, which yields the higher value $K \approx 3.8$.
The $K$ value of a dense, uniform, and symmetric cluster is reached for $\theta \approx \pi/3$ ((Fig.\ref{Screenshot2}(b)). The larger $K$ value at $\pi/2$ is caused by the shape asymmetry of the cluster in in the HCP phase (Fig.\ref{Screenshot2}(a)). 
From the kurtosis it is clear that the dilute phase and closed packed hexagonal structure phases are more uniformly distributed and the value is close to the theoretical expectation.
Hence, the kurtosis allows for a distinction of the dilute, aggregate (or worm-aggregate), and HCP phase and shape asymmetry within a phase.

\subsubsection{Active diffusion coefficient}

\begin{figure}
\includegraphics[width = \columnwidth]{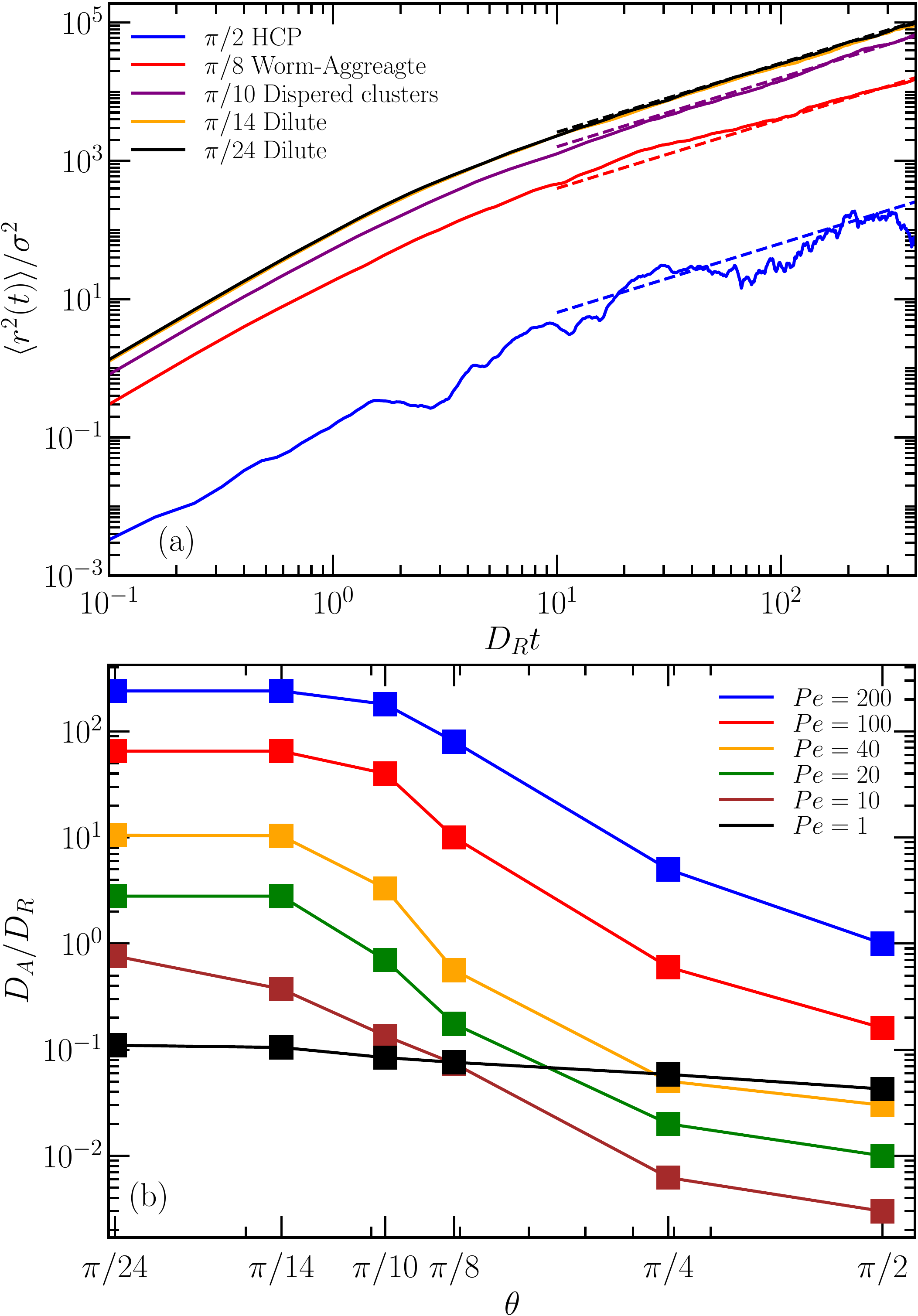}
\caption{(a) Mean square displacement of ABPs as a function of time for $Pe=100$. The dashed lines are fits to a linear time dependence. (b) Effective diffusion coefficients, $D_A$,  as a function of the vision angle $\theta$ for various $Pe$. The other parameters are  $\Omega=5$, $R_0=1.5\sigma$, and  $\Phi = 0.0785$. }
\label{Fig:MSD_low}
\end{figure}

Figure \ref{Fig:MSD_low}(a) provides examples of iABP mean-square displacements. From the linear regime for $t D_R >1$, we extract the effective active diffusion coefficient $D_A$ according to $\langle \bm r^{2}(t) \rangle =4D_A t$. The respective values $D_A$ are presented in Fig.~\ref{Fig:MSD_low}(b). 
For the vision angles $\theta < \pi/14$ --- the dilute regime --- the iABPs behave as ABPs in dilute solution, and $D_A = D_T + v_0^2/(2D_R)$. The values from our simulations are  within $20-30\%$ of the expected theoretical values, despite the additional vision interaction. With increasing $\theta$, aggregates form and  $D_A$ decreases. Even for P\'eclet numbers as small as $Pe=1$, we observe a decrease of $D_A$, although small compared to that at larger $Pe$, suggesting emergent inhomogeneities. For all larger $Pe$, we obtain a very similar trend with increasing $\theta$. At $\theta = \pi/2$, the iABPs are in the HCP phase for $Pe \geq 10$. Here, the iABPs diffuse together as a cluster, with an diffusion coefficient $D_A \sim 1/N$, a relation predicted for ABPs in an aggregate (cf. Sec.~\ref{sec:msd_low}.\cite{eise:16}

\section{Cluster growth} \label{sec:cluster_growth}

In the HCP phase, clusters form from an initial uniform distribution of iABPs, which then merge and grow in time.  We analyze the cluster-growth process as a function of time by calculate the number $p(n,t)$ of clusters of size $n$  at time $t$ and determine the time-dependent average cluster size  \cite{evans1999colloidal}
\begin{equation}
    C(t) = \frac{1}{N_c(t)}\sum_{n = 1}^{N_c(t)}p(n,t) ,
\end{equation}
where $N_c(t)$ is total number of cluster at time $t$. As before, particles belong to a clusters if the distance between  ABPs is smaller than $\sigma_0 = 1.2\sigma$. For convenience,  we consider the larger packing fraction $\Phi = 0.157$, with a  phase diagram very similar to that in Fig.~\ref{Fig:Phase_Pe_vs_theta_high_den} up to $Pe \approx 20$, only the boundaries of the dilute and aggregate phase  are somewhat shifted (Figs. S2). In particular, we observe a HPC phase for $1< Pe <20$ at $\theta = \pi/2$. 

\begin{figure}[t]
\begin{center}
\includegraphics[width =\columnwidth]{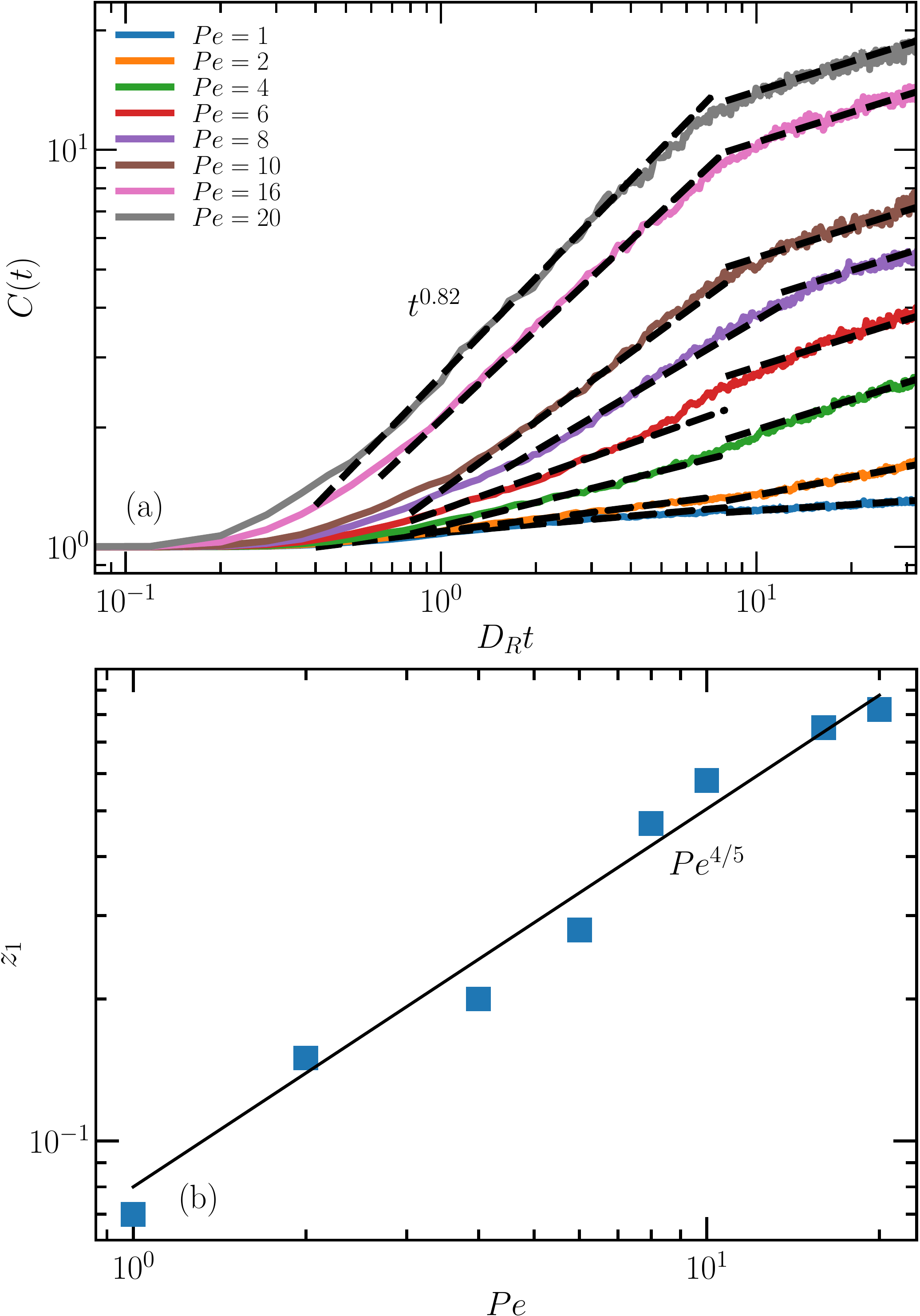} 
\caption{(a) Average cluster size as a function of time for various $Pe$, the vision angle $\theta=\pi/2$, and the packing fraction $\Phi = 1.57\times 10^{-1}$. (b) Growth exponent $z_1$ as a function of $Pe$. See movie M6 (ESI$\dag$).
}
\label{Cluster_growth}
\end{center}
\end{figure}

%%%%%%%%%%%%%%%%%%%%%%%%%%%%%%%%%%%%%%%%%%%%%%%%%%%%%%%%%%%%%%%%%%%%%%%%%%%%%%%%%%%%%

Figure~\ref{Cluster_growth}(a) shows the time-dependence of the average cluster size for various $Pe$. Evidently, cluster growth depends on $Pe$, and $C(t)$ exhibits two power-law regimes, $C(t) \sim t^{z_k}  (k=1,2)$. The short-time regime, $1 < D_Rt <10$, starts approximately after the iABPs reach the active diffusive time regime (Fig.~\ref{Fig:MSD_low}).  The exponent of the power-law growth, $z_1$, depends on $Pe$. As displayed in Fig.~\ref{Cluster_growth}(b), it can  be well described by $z_1 \sim Pe^{4/5}$. Even more, the exponent depends on the vision angle (Figs. S5) for which we find $z_1 \sim \theta^{1/5}$, hence,  
\begin{equation} \label{eq:exponent_m}
      z_1 \sim Pe^{4/5} \theta^{1/5}  .
\end{equation}
For $D_Rt \gtrsim 8$, $C(t)$ crosses over to a second power-law regime, where $C(t) =Q(Pe,\theta)  t^{z_2}$, with $z_2=1/4$ independent of $Pe$, but the prefactor, $Q$, is $Pe$ dependent. Similarly, at constant $Pe$, $z_2$ is independent of $\theta$, but the prefactor is $\theta$ dependent. However, this applies  for $Pe \geq 4$ and $\theta \geq \pi/6$ only.  Both time regimes are determined by the maneuverability  of the iABPs.

%%%%%%%%%%%%%%%%%%%%%%%%%%%%%%%%%%%%%%%%%%%%%%%%%%%%%%%%%%%%%%%%%%%%%%%%%%%%5555
\begin{figure}[t]
\includegraphics[width = \columnwidth]{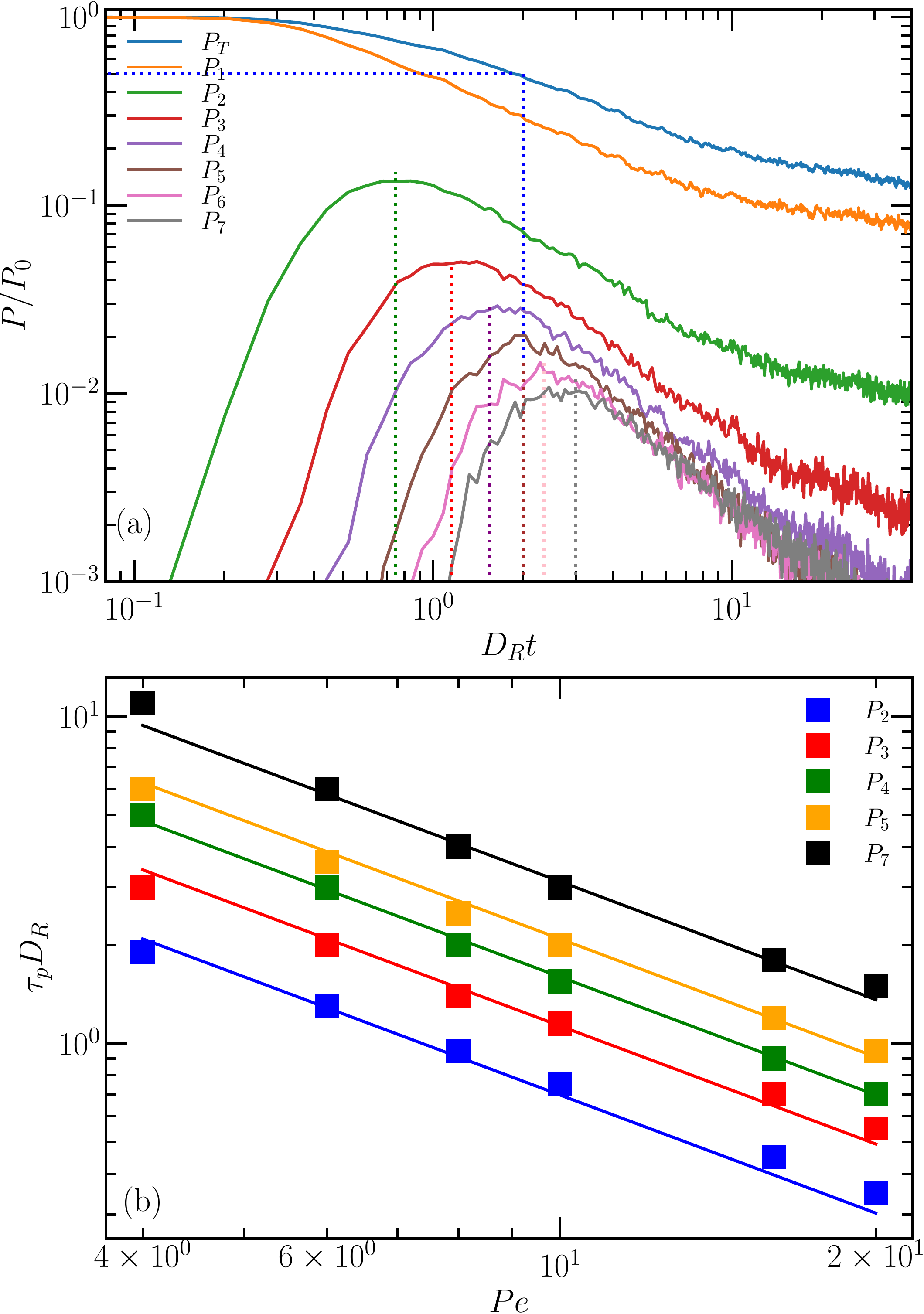}
\caption{(a) Temporal evolution of the cluster concentrations $P_i$ as a function of time for the vision angle $\theta=\pi/2$ and activity $Pe=10$. $P_T$ is the total concentration. (b) Characteristic times of the maxima of $P_i(t)$ as a function of $Pe$. The fitted line corresponds to  equation \ref{eq:maximum_time}, whereas the squares represents simulation results. }
\label{Probablity_evolution}
\end{figure}
%%%%%%%%%%%%%%%%%%%%%%%%%%%%%%%%%%%%%%%%%%%%%%%%%%%%%%%%%%%%%%%%%%%%%%%%%%%%%%%%5

To further characterize cluster formation, we consider the concentration $P_n(t)= p(n,t)/A$ (Eq.~\ref{eq:cluster}) of clusters containing $n$ iABPs at time $t$ within the area $A$ of the simulation box. Initially the iABP concentration is $P_0=0.2$. The temporal evolution of the relative concentration of clusters is presented in Fig.~\ref{Probablity_evolution}(a) for various clusters sizes. The concentration  $P_1$ of individual iABPs decrease with increasing time as particles merge. At the same time, clusters emerge, first dimers, then trimers, etc., into larger clusters, which causes an initial growth of $P_n(t)$ ($n>1$), reaching a maximum, and a decrease  with increasing time. 
%calculations for short-range attractive passive colloids predict the time dependence $P(i,t) = P_0 (t/\tau_0)^{i-1} (1+t/\tau_0)^{-i-1}$, and $\tau_0$ a characteristic time determined by the rate constant \cite{evans1999colloidal}. In the limit $t/\tau_0 \gg 1$, the asymptotic behavior $p(n,t) \sim 1/t^2$ is obtained, independent of $n$. In contrast, our simulations yield $P(n,t) \sim 1/t$, hence, the clustering dynamics is different from that of passive short-range attractive colloids. 

The characteristic time $\tau_p(n)$, where $P_n(t)$ assumes its maximum, is presented in Fig.~\ref{Probablity_evolution}(b) for various cluster sizes and $Pe$. The time $\tau_p$ increase with increasing $n$ and decreases with increasing $Pe$, where the data sets are well described by the relation $\tau_p D_R \sim (n/Pe)^{\kappa}$, with $\kappa=6/5$. In addition, $\tau_p$ depends on and vision angle $\theta$ (Supplementary Material). Overall, the maximum time is well described by the expression  
\begin{equation} \label{eq:maximum_time}
\tau_p(n) D_R=5.5 (n/Pe)^{\kappa}\theta^{\kappa/4} . 
\end{equation}

Theoretical calculations for short-range attractive passive colloids based on the Smoluchowski equation with a constant kernel (cluster-size-independent rate constant) predict the asymptotic behavior $P(n,t) \sim 1/t^2$ for $t\to \infty$.\cite{evans1999colloidal} Moreover, the time at the maximum of $P(n,t=\tau_p)$, $\tau_p \sim n$, increases linearly with $n$. This is distinct from our simulations, with $P(n,t) \sim 1/t$ over a certain time regime, at least for $n>2$, and the super-linear increase of $\tau_p$ of Eq.~\eqref{eq:maximum_time}.  Hence,  the clustering dynamics with activity differs strongly from that of passive short-range attractive colloids.

%Fig. \ref{Probablity_evolution} (d) represents the variation of half life time $\tau$  with $Pe$ at fixed vision angle $\pi/2$. Half life time is time at which the total concentration is reduced to half the initial value. The dependence in $Pe$ is given as $ \tau \sim Pe^{-9/5}$. For higher activities the ABPs are faster so easily travel more and form clusters faster so half life times is shorter.

\section{Summary and Conclusions}

We have studied structure formation in systems of intelligent active Brownian particles, adopting the cognitive flocking  model of Ref. [\onlinecite{barb:16}]. We systematically varied the maneuverability strength, the vision angle and vision cutoff radius, and the ABP activity. We find distinct phases, such as a dilute fluid,  aggregates, worms, worm-aggregate coexistence, hexagonally closed packed structures (HCP), and a diffuse clusters phase,  depending upon the above parameters. We established phase diagrams for various packing fractions, illustrating the emergent phases. 

Larger vision angles, $\theta$, lead to a larger number of sensed particles and a stronger response. Correspondingly, the system is in a dilute, fluid-like phase at lower $\theta$, and at large angles dense, hexagonally close-packed structures appear. In-between, dependent on the maneuverability strength, additional phases emerge, which have not been found in previous studies, specifically the coexistence of worms and aggregates.

The analysis of the iABP dynamics shows, in the dilute and worm phase, mean-square displacements identical with those of an individual ABP for the considered packing fractions  $\Phi \le 0.16$. The long-time active diffusion coefficient, $D_A$, decreases with increasing vision angle, and in the HCP phase, $D_A \sim 1/N$, where $N$ is the number of particles in the cluster. Here, the iABP move together in a diffusive manner.

Starting from a homogeneous and isotropic system for the (large) vision angle $\theta = \pi/2$, the iABPs begin to nucleate and to form clusters, which merge in the course of time. Our analysis of the temporal evolution of the average  cluster size, $C(t) \sim t^{z_k}$, reveals two power-law regimes with characteristic short-, $z_1$, and long-time, $z_2$, exponents. The exponent $z_1 \sim Pe^{4/5} \theta^{1/5}$ strongly depends on activity and the vision angle, whereas  $z_2=1/4$ is independent of $Pe$. The analysis of the temporal evolution of the cluster concentration, $P_n$, yields the power-law dependence $\tau_p(n) D_R = 5.5(n/Pe)^\kappa \theta^{\kappa/4}$, with $\kappa=6/5$ for the characteristic time, where $P_n$ assumes a maximum. The exponent $z$ is larger than that obtain in calculations of short-range attractive passive colloids, where $\tau_p \sim  n$, or $\kappa=1$.\cite{evans1999colloidal} Hence, the implemented perception rule slows down clustering and cohesion at large clusters. Our considerations show that active processes can lead to an aggregation process, which is distinct from that of passive systems.

Moreover, aggregation depends on the underlying active process. The considerations in Ref.~[\onlinecite{cremer2014}] for ballistic and diffusive active motion of clusters yield the time dependencies $t^1$ and $t^{1/2}$, respectively, for cluster growth in two dimensions at long times. Compared to these, the exponent $z_2=1/4$ suggests a much slower cluster growth for the considered vision-based active system. 

Inertia effects can play a significant role in the structure formation of active systems.\cite{das:19,loew:20} This is also reflected in our simulations with  underdamped equations of motion. Despite a large friction coefficient and a crossover in the mean-square displacement from the short-time passive ballistic motion to the over-damped diffusive motion at $\langle \bm r^2 \rangle /\sigma^2 \approx 10^{-3}$, a fraction of a colloid diameter, the phase boundary are somewhat shifted, compared to an overdamped dynamics.          

We have considered a minimal model of iABPs, with vision-based input and limited maneuverability, which  serves as a basis for various possible future extensions to more specific or advanced cognitive processes. In particular, multiple input channels complemented by a complex decision process and response can be designed, for which we expect other collective dynamical features and means to control the emerging structures and collective dynamics.

%\section*{Author Contributions}
%We strongly encourage authors to include author contributions and recommend using %\href{https://casrai.org/credit/}{CRediT} for standardised contribution descriptions. Please refer to our %%general \href{https://www.rsc.org/journals-books-databases/journal-authors-reviewers/author-responsibilities/}%{author guidelines} for more information about authorship.

\section*{Conflicts of interest}
There is no conflict of interest .

%\section*{Acknowledgements}
%The Acknowledgements come at the end of an article after Conflicts of interest and before the Notes and %references.

%%%END OF MAIN TEXT%%%



%%%REFERENCES%%%

%\bibliography{bibliography.bib,CollectiveBABP.bib} 

%\bibliography{/Users/winkler/ownCloud/publications_library/bibliography/bibliography.bib,CollectiveBABP.bib} 

%\bibliographystyle{rsc} %the RSC's .bst file

\providecommand*{\mcitethebibliography}{\thebibliography}
\csname @ifundefined\endcsname{endmcitethebibliography}
{\let\endmcitethebibliography\endthebibliography}{}

\end{document}